\documentclass[11pt]{article}

\usepackage{amsmath}
\usepackage{enumerate}
\def\be{\begin{equation}}
\def\bea{\begin{eqnarray}}
\def\ee{\end{equation}}
\def\eea{\end{eqnarray}}
\def\d{\partial}
\def\eps{\varepsilon}
\def\la{\lambda}

\def\be{\begin{equation}}
\def\beaa{\begin{eqnarray}}
\def\ee{\end{equation}}
\def\eeaa{\end{eqnarray}}
\def \bea {\be}
\def \eea{\ee}

\def\vph{\varphi}

\def\k{\kappa}
\def\ov{\over}

\def \bi{\bibitem}
\def \lab {\label}

\def \l {\lambda}
\def\foot{\footnote}
\def \adss {$AdS_5 \times S^5~$ }
\newcommand{\rf}[1]{(\ref{#1})}
\def \ov {\over}
\def \tr {{\rm tr}}

\def \td {\tilde}
\def \ci{\cite}

\def\id{\protect{{1 \kern-.28em {\rm l}}}}

\def \no {\nonumber} 
\begin{document}

\overfullrule=0pt
\parskip=2pt
\parindent=12pt
\headheight=0in \headsep=0in \topmargin=0in \oddsidemargin=0in

\vspace{ -3cm}
\thispagestyle{empty}
\vspace{-1cm}

\today

\rightline{Imperial-TP-AT-2014-07}
\rightline{   }

\begin{center}

\vspace{1cm}
{\Large\bf
%Target space
Supergravity  backgrounds for  deformations \\
\vspace{0.2cm}
of AdS$_n \times $S$^n$ supercoset string models 
}
\vspace{1.4cm}

{O. Lunin$^{a,}$\footnote{olunin@albany.edu  }, R. Roiban$^{b,c,}$\footnote{radu@phys.psu.edu} \ and A.A. Tseytlin$^{d,}$\footnote{Also at Lebedev Institute, Moscow. tseytlin@imperial.ac.uk }}\\

\vskip 0.3cm

{\em $^{a}$    Department of Physics,
University at Albany (SUNY),
Albany, NY 12222, USA
 }
\vskip 0.08cm

{\em $^{b}$Department of Physics, The Pennsylvania State University,\\
University Park, PA 16802 , USA}
\vskip 0.08cm

{\em $^{c}$Kavli Institute for Theoretical Physics,
University of California,\\
 Santa Barbara CA 93106,  USA}
\vskip 0.08cm

{\em $^{d}$ The Blackett Laboratory, Imperial College, London SW7 2AZ, U.K.}

\vspace{.2cm}

\end{center}

{\baselineskip 11pt
\begin{abstract}
\noindent
%AT
We consider  type IIB supergravity   backgrounds 
 corresponding to  the deformed AdS$_n \times$ S$^n \times $T$^{10-2n}$ supercoset  string models 
 of the type constructed in    arXiv:1309.5850  which depend on one deformation parameter $\k$.
In  AdS$_2 \times$ S$^2$  case we   find  that the deformed metric  can be   extended to a 
 full  supergravity  solution   with non-trivial dilaton, RR scalar and RR 5-form  strength. %(or a vector in 4d). 
 The solution  depends   on a  free parameter $a$ that  should  be chosen  as a  particular  function of $\k$
 to  correspond  to the deformed supercoset model. 
In  AdS$_3 \times$ S$^3$  case the full solution supported by  the dilaton, RR scalar and  RR 3-form   strength 
exists only in the two special cases of $a=0$ and $a=1$. We conjecture that there may be   a more general one-parameter solution 
 supported by several RR  fields    that  for particular $a=a(\k)$  corresponds to the supercoset model. 
In the  most complicated   deformed AdS$_5 \times$S$^5$  case we   were able  to find  only
the expressions  for the dilaton  and the RR scalar. % consistent with the form of the deformed  metric. 
 The full solution is likely to be supported
   by a combination of the  5-form and 3-form  field strengths. 
   We comment on the singularity structure of the resulting  metric and exact dilaton field. 
\end{abstract}
}

%\maketitle

\newpage

\def \l {\lambda} 
\def \ba {\begin{align}}
\def \ea {\end{align}}
\def \lab   {\label}  \def \r {\rho}
\def \adst {AdS$_2 \times$S$^2$\ }
\def \la {\label}

\def \iffa {\iffalse}
\def \adstr {AdS$_3 \times$S$^3$\ }
\def \adss {AdS$_5 \times$S$^5$\ }

\def \z {\zeta}
\def \a {\alpha}\def \del {\partial}

\def \ed {\end{document}}
\def \adn {AdS$_n \times$S$^n$\ }
\def \we {\wedge}
%%%%%%%%%%%%%%%%%%%%%%%%%%%%%%%%%%%%%%
\tableofcontents
%%%%%%%%%%%%%%%%%%%%%%%%%%%%%%%%%%%%%%

\date{\today}

\setcounter{footnote}{0}
\setcounter{section}{0}

%\newpage 
%%%%%%%%%%%%%%%%%%%%%%%%%%%
%\newpage 
\section{Introduction}
%%%%%%%%%%%%%%%%%%%%%%%%%%%%%%%

Integrability of string sigma model is a key feature that  allows  to determine 
 the string spectrum in non-trivial curved backgrounds. The study of integrable deformations  of the
  most-symmetric \adss  model underlying AdS/CFT correspondence 
\ci{beir}  is thus an important avenue of research that may also shed light  on  hidden symmetries of dual gauge theories.  
Recently, a novel  one-parameter
  integrable deformation of  the 
\adss  supercoset model  was  constructed in \ci{DMV}  (see  also \ci{ABF,HRT,Arutynov:2014ota,Arutyunov:2014cra,Delduc:2014kha,Hollowood:2014rla}). This model  generalizes some  previously known low-dimensional  bosonic  integrable models 
\ci{foz,fat,k2,luk,k3}.  

The corresponding  target space  type IIB supergravity background  
 has   no space-time   supersymmetry and the  bosonic isometry is reduced from 
 $SO(2,4) \times SO(6)$ to its  Cartan   subgroup $[SO(2)]^3 \times  [SO(2)]^3$, i.e.  most of the symmetry of  the original  \adss  space 
 becomes hidden  (or ``$q$-deformed"). 
Starting with a specific parametrization of the  bosonic part  of  the 
deformed  supercoset model \ci{DMV} the corresponding  10d  metric  and $B$-field were   found explicitly  in \ci{ABF}. However, 
extracting  the associated RR field  strengths  that should 
 promote  the deformed metric to an exact supergravity solution %(as suggested by kappa-invariance of the
% deformed supercoset   action \ci{DMV})
from the fermionic part of the supercoset action 
turns out to be challenging even in the 
simpler  low-dimensional  \adst and \adstr  models \ci{HRT}.

Our  aim here will be  to find the deformed  \adn  type IIB  backgrounds  by  (i) starting with 
the deformed   metric  as given by the bosonic part of the supercoset model  and 
(ii) solving  the  supergravity equations   directly to find  the expressions of all other fields  required to 
support  this metric as  an exact   solution. 
Finding    ``matter"  fields   supporting a given metric via Einstein equations 
 is  not a standard   GR problem;
the solution may not exist or, if it exists, it may not be unique.
%there may be  many  solutions.  
The  present case  is complicated also   by the  absence   of  supersymmetry and   non-abelian  isometries. 
We shall   see   that the solutions will have a rather unusual feature: 
while the  string-frame  metric   is  a direct sum of the  deformed AdS$_n$  and S$^n$ parts,  this will  no longer  be so  
for the dilaton and the RR fields -- they  will  not factorize and thus ``tie"   the   AdS$_n$  and S$^n$ parts   together 
(as  what fermion part  of supercoset model does). 

Having  found a  supergravity solution  with the required   deformed \adn  metric,  one is still to decide 
if  it is the one that actually corresponds  to the integrable  deformed  supercoset model of \ci{DMV}. 
As we shall see below, the  solution for the dilaton and RR fluxes supporting a given  deformed metric is 
not unique:  in \adst  case there is a one-parameter $a$-family  of solutions,  and the same  is  expected to be the case also 
in the \adstr and \adss cases.  One is then to   choose $a$  as a function  of the 
 the deformation parameter $\k$ \foot{We shall follow \ci{ABF}  and 
 use $\k= {2 \eta \ov 1-\eta^2}$  as  the deformation parameter, where $\eta$ is   the parameter used in  \ci{DMV}.}
 in order to match the  supercoset model. 
This  choice   may    be aided by  consideration of the two special limits  discussed in \ci{HRT}:
\begin{enumerate}[(i)]
\item $\k =\infty$  or  ``maximal deformation limit":  in this  case    the  deformed \adn  supercoset model becomes  T-dual to
 ``double Wick rotation"  of the 
undeformed \adn  model, i.e. it has dS$_n \times $H$^n$   target space  supported by an imaginary $n$-form  RR  flux; 
\item
$\k= i$  (combined with a rescaling of coordinates and string tension) or  ``pp-wave limit":
 in this case    the target-space   metric becomes of pp-wave type  and the problem of 
  finding the  supporting dilaton and fluxes  simplifies.
\end{enumerate}

We shall start  in section 2  by   finding a  one-parameter  type IIB solution
with the metric being  that 
 of the $\k$-deformation of the  AdS$_2\times$S$^2 \times $T$^6$   one    \cite{HRT}. %  to a 10d  type IIB supergravity solution. 
It  corresponds to 
 a  solution of  4d supergravity obtained by compactification on 6-torus  with  only  the 
 dilaton, RR scalar  and the RR 2-form being  non-trivial. 
 Guided by the two special   limits  mentioned above  we  shall  argue that  for a special  value  of the free parameter 
 $a=a(\k) =  \k^{-1} \eta 
  = \k^{-2} ( \sqrt{ \k^2 + 1} - 1)$    the resulting   background  should 
   corresponds to the $\kappa$-deformation of the AdS$_2\times$S$^2$ supercoset model.
   
In section 3  we shall    consider 
 the $\kappa$-deformation of the AdS$_3 \times $S$^3\times $T$^4$
space supported by the  RR 3-form flux.  Compactifying on 4-torus we shall use  the  truncated   6d action 
containing the metric, dilaton, RR scalar and RR 3-form field.
  Starting with the $\k$-deformed \adstr   metric \ci{HRT} we   will 
find  again  a one-parameter family of solutions of  the three scalar equations. However,   only 
 two special  members of this family  (with $a=0$ and $a=1$) will  have  extensions 
  to solutions of  the full set of 6d  supergravity equations if only one RR 3-form field is assumed to be non-zero. 
  The existence of   the  complete   solution   with an arbitrary  parameter $a$  (that  may  be  chosen again as $a(\k)$ to match the 
  deformed supercoset model)  appears  to  require  more  RR    field strengths to be non-zero, a possibility which remains to be studied.  %   and remains to be  found explicitly. 
  %AT
  We shall  also  present the  analogs of the $a=0$ and $a=1$   solutions in the case   of 2-parameter  ($\k_+,\k_-)$  deformation of the 
  \adstr supercoset \ci{ben}   with the  metric corresponding to the  2-parameter Fateev model  \ci{fat} for deformations of 
  AdS$_3$ and S$^3$.

Guided   by  the low-dimensional examples,   in section 5 we shall  address  the problem of 
 promoting  the $\k$-deformed \adss   metric and the  $B$-field   found in \ci{ABF}  to the  full type IIB supergravity solution.
 An additional complication  is  that the 10d metric (and thus  also other  background fields) 
    contains a   non-trivial dependence on two extra  angular coordinates. 
 We will  present  two  special  solutions to the    equations  for the dilaton and the RR scalar 
  which are the counterparts of 
 the $a=0$ and $a=1$  solutions in the AdS$_3\times$S$^3$ case. 
 Here we will    not able  to  find   the corresponding 5-form flux  and it  appears likely that the full solution   should  exist
 only  when  also the  RR 3-form  flux is non-zero. 
 
     Some comments on the singularity  properties of   the deformed \adn backgrounds  will   be  included in  section 5.
 In appendix A   we will give the form of the relevant supergravity equations in  different dimensions
 %v2
 and discuss truncations of  the 10d supergravity  action.
 In appendix B we  will  review 
 the algebraic Rainich conditions on Maxwell  stress tensor in 4 dimensions.

%%%%%%%%%%%%%%%%%%%%%%%%%%%%%%%%%%%%%%%%%%%%%%%%%%%%%%
\renewcommand{\theequation}{2.\arabic{equation}}
\setcounter{equation}{0}

\section{Deformation of AdS$_2\times$S$^2$}
\label{Sec4D}

In this section we  shall extend the  metric
 of the deformation of the  AdS$_2\times$S$^2 \times $T$^6$    space 
   \cite{HRT}  to a 10d  type IIB supergravity solution. 
 Similarly to the undeformed background \ci{STWZ},  this   solution is a direct 10d lift 
 of the corresponding  solution of 4d supergravity obtained by compactification on 6-torus: 
 the 5-form field strength $F_5$  is  given by the  product of the non-trivial 2-form field   strength $F_2$ 
  in 4 dimensions  and the  canonical  hermitian 3-form of T$^6$.\foot{
  This  background  can be embedded into type II supergravity  as described in Appendix A, see 
   (\ref{Metr10D6red}), (\ref{StarF1110})  and  (\ref{EmbedIIB}).
%  Explicitly, 
%  $F_5 = F_2 \we \Omega_3 + \star$, 
%  \ \ \  $ \Omega_3 (T^6)=  {\rm Re} ( dz_1 \we dz_2 \we dz_3)$.
}
   The background fields (metric, dilaton, RR scalar and 1-form potential) will  depend  on a free parameter $a$. % that can be chosen  as a special function $a(\k)$ 
We shall conjecture that  for a special choice   of $a= a(\k)$  the resulting   background should 
   correspond to the  superstring sigma model which is   the 
   $\kappa$-deformation of the AdS$_2\times$S$^2$ supercoset model  based on 
PSU$(1,1|2)$/U$(1)\times$U$(1)$.
%
%This solution will have a nontrivial dilaton and RR scalar and a RR two-form field strength which descends from the ten-dimensional five-form 
%field strength. 
%The background fields will  depend  on a free parameter $a$ that can be chosen  as a special function
As a check, we  shall  show  that in the   special limits      of  $\k= \infty $ ($a=0$) and $\k=i$ ($a=1$)
we  indeed reproduce  the   expressions  expected from the deformed  supercoset  construction. 

 %match the known special  limits for 
%subsequently fixed to a specific function of $\kappa$ such 
%that limits of the supergravity field configuration reproduce the results of the coset analysis.

%%%%%%%%%%%%%%%%%%%%%%%%%%%%%%%%%%%%
\subsection{One--parameter family of  solutions}
\label{SubAdS2soln}

Let us recall 
 that the AdS$_2\times$S$^2 \times $T$^6$ solution can be obtained as  a limit of 10d   type IIB   solution describing 
four intersecting stacks  of D3-branes  (see, e.g., \cite{STWZ} and refs. there). 
Upon reduction on $T^6$ to four dimensions it is supported by a two-form field strength $F_2$. 
The compactification of type II supergravity to four dimensions on T$^6$  in general 
contains a large number of scalar and vector  fields, some of which describe the deformations of the  compact space. 
Since by   construction 
 the deformation  acts only on the supercoset part of the geometry we may 
%expect that the compact space is not deformed. With this assumption, the other four-dimensional fields that can be 
assume that the fields that   should be  non-vanishing  are   not related to  T$^6$. 
The minimal choice is the metric,  dilaton,    the 
 RR  scalar $C$  and   the  vector $A$   (with    $F_2=dA$ as  its field strength; 
the latter  may represent several identified components of the 10d fields, 
%v2
cf. Appendix A). 
The  Lagrangian  for 4d  supergravity restricted to these fields is\foot{In general, 
the action may   contain also a term $ \a CF_{mn}{\tilde F}^{mn}$  with some special constant $\a$. 
It is possible to show that in the present case  one should choose  the 
 identification of the fields  such that $\alpha=0$ as 
otherwise   one  will   not get the   undeformed \adst  background as a solution.}
\bea
{\cal L}_4=e^{-2\Phi}\big[ R+4(\nabla\Phi)^2\big] -
\frac{1}{4}F_{mn}F^{mn}-\frac{1}{2}(\del C)^2 \ . 
%+\alpha CF_{mn}{\tilde F}^{mn} \ .
\label{4daction}
\eea
The simplest  solution is the \adst Bertotti-Robinson one   with $\Phi$ and $C$ being constant   and 
\ba \label{2233}
&ds^2={L^2}
\Big[-(1+\rho^2)dt^2+\frac{d\rho^2}{1+\rho^2}\Big]+
{L^2}
\Big[(1-r^2)d\vph^2+\frac{dr^2}{1-r^2}\Big] \  , \\
& F_2= 2 L ( c_1  d\rho\wedge dt+ c_2 dr\wedge d\vph) \ , \ \ \ \ \  \ \ \ \ \ \ \ \   c_1^2 + c_2^2=1 \ . \la{3322}
\end{align} 
Here $c_1,c_2 $  are   reflecting the  freedom of  $U(1)$ 
electromagnetic duality  rotations. %, with  the symmetric   choice being $c_1=c_2={1\ov \sqrt 2} $. 
%$L$ is the overall  curvature scale  that we shall often set to 1 in what follows.  

%Here $\a$ is a constant that  we should set to be zero in what  and 
Our  aim will be to find $\Phi, C$ and $ F_2$   that   promote 
the deformed AdS$_2\times$S$^2$  metric   \cite{HRT} 
\be\label{22}
ds^2=\frac{L^2}{1-\kappa^2\rho^2}
\Big[-(1+\rho^2)dt^2+\frac{d\rho^2}{1+\rho^2}\Big]+
\frac{L^2}{1+\kappa^2r^2}
\Big[(1-r^2)d\vph^2+\frac{dr^2}{1-r^2}\Big] \ 
\ee
to  an exact solution of the theory \rf{4daction}. 
Here $L$ is the  (inverse) curvature  scale   (that we shall  often set to 1 in what follows) and $\k$  is the  parameter of  deformation
 away from the symmetric AdS$_2\times$S$^2$  point.
The  Ricci tensor    and the curvature scalar of the metric $g^A \oplus g^S$ in  \rf{22}  can be written as
\ba \la{uc} 
&R^A_{ab}=-(1 + \k^2)\frac{1 + \k^2\rho^2}{1 - \k^2\rho^2}
\ g^A_{ab},\qquad\qquad 
R^S_{ab}=(1 + \k^2)\frac{1 - \k^2r^2}{1 + \k^2r^2}\ g^S_{ab}
\\ \la{cu}
&R= 4   ( 1 + \k^2)  \Big( - {1 \ov 1 - \k^2 \r^2} + {1 \ov 1 + \k^2 r^2}  \Big) \ . \end{align}
%
%
%%%%%%%%%%%%%%%%%%%%%
%
%
%In appendix \ref{SecExtraCoupling} we will look at a particular limit of the metric (\ref{22}) to demonstrate that $\alpha=0$, and here we only 
%focus on this case. 
The equations of motion following from  \eqref{4daction} are  given  in the appendix \ref{AppEOM};  we shall focus first
on  the trace of the Einstein equation, the equation for the RR scalar  and  the equation for the dilaton that can be organized as (cf. \eqref{cc})  
%Slightly rearranging them yields:
\bea
R+ 
2 \nabla^2\Phi  + \frac{1}{2}e^{ 2\Phi}\d_m C\d^m C=0 \ , \qquad \nabla^2 C=0\ , \qquad 
\label{23}
\nabla^2(C^2+4e^{-2\Phi})=0 \ .
\eea
A way to solve this  system  is to consider  first 
a particular  limit: a   small $\k$ expansion  combined with a particular  rescaling of the coordinates 
%that makes  $S^2$  metric  undeformed  but keeps  the  deformation of AdS$_2$ non-trivial 
\bea \la{24} 
\k\rightarrow 0\ ,\qquad  \mbox{with \ fixed} \qquad  \k \r\ ,\ \ \
\k^{-1}  t  \ , 
%\eps\rightarrow 0\ ,\qquad  \mbox{with \ fixed} \qquad \quad {\tilde \k}=\frac{\k}{\eps},\quad
%{\tilde \rho}=\eps\rho,\quad {\tilde t}=\frac{t}{\eps}  \ \ , 
\eea
in which the $S^2$   part of the metric  becomes undeformed  while the deformation of the  AdS$_2$  part  remains non-trivial, i.e.
 %we fist notice that it reduces to ordinary differential equations in the limit 
%in (\ref{22}). 
%Dropping tildes, we find the metric
%Dropping tildes  we get 
\bea\label{25}
ds^2=\frac{1}{1- ( \k \rho)^2}
\Big[-\rho^2 dt^2+\frac{d\rho^2}{\rho^2}\Big]+
\Big[(1-r^2)d\vph^2+\frac{dr^2}{1-r^2}\Big] \ . 
\eea
%which is symmetric under the rescaling 
%\bea
%\k\rightarrow \la \k,\quad \rho\rightarrow \frac{\rho}{\la},\quad t\rightarrow \la t. \eea
%To respect this symmetry,
 The perturbative expansion in $\k$    respecting  the symmetry $ \k\rightarrow \l \k,\ \rho\rightarrow \frac{\rho}{\l},\ t\rightarrow \l t$
 of the metric \rf{25} 
  should then be an expansion in  powers of $\k \r$:
\bea\label{PertExpDil}
e^{-\Phi}=1+\sum ( \k \rho)^n f_n(r),\qquad
C=\sum ( \k \rho)^n g_n(r).
\eea
Substituting this into (\ref{23}) and summing up the perturbative series, we find the most general solution
corresponding to the   metric \rf{25}. The  solution  depends on one  free parameter $a$: 
\ba
%\label{27}
 e^{-2\Phi} =\frac{(1 - \k^2\rho^2)}{1-  a^2 ( \k \rho)^2+ ( \k \rho r)^2 -2\k\sqrt{1-a^2} \ \rho r}\ , \qquad  C= 2\sqrt{\frac{1}{a^2}-e^{-2\Phi}} \ . \lab{28}
\end{align}
%which depends on one parameter $a$. 
Note that   here the combination $C^2+4e^{-2\Phi}$   that  should be  a harmonic function  according to  \rf{23}   is simply a  constant
\bea\label{29}
C^2+4e^{-2\Phi}=\frac{4}{a^2}  \ . 
\eea
%remains constant.
%must remain constant, and we will {\it impose} this requirement on the solution supporting the full metric (\ref{22}). 
%This leads to the unique solution for the scalar fields, and the other fluxes can be found as well:
Going back to the general case of the metric  \rf{22}  and 
requiring that the   solution of \eqref{23}  should have  the same property \rf{29} 
 leads to a  similar  one-parameter solution for the 
scalar fields. It is then   easy to find also the   solution for  the vector  potential $A$\ \footnote{One may check that  the candidate $F_2$   implied  
 by the form of the Maxwell 
stress tensor appearing  on the right-hand side of the 
 Einstein's  equations   for the given metric  and the scalar fields   obeys the algebraic Rainich condition 
\cite{Rainich_original,Misner:1957mt,Torre:2012hw,Torre:2013nia} (see appendix B), i.e.   there 
should indeed exist a  vector  field sourcing this geometry.} 
%The two-form flux can be found as well:
\ba %\label{299}
ds^2&=\frac{1}{1 - \k^2\rho^2}
\Big[-(1+\rho^2)dt^2+\frac{d\rho^2}{1+\rho^2}\Big]+
\frac{1}{1 + \k^2r^2}
\Big[(1-r^2)d\vph^2+\frac{dr^2}{1-r^2}\Big] \ , \la{199}\\
e^{-2\Phi} &= %\frac{(1 - \k^2\rho^2) (1 + \k^2 r^2)}{ }\equiv
\frac{(1 - \k^2\rho^2) (1 + \k^2 r^2)}{P(\r,r)}, \ \ \ \ \  \ \ \      P(\r,r) \equiv 1 + \k^2\big[a^2 (r^2-\rho^2)-2b\, r\rho+r^2\rho^2\big]\ , 
\la{999}\\
C&=2\sqrt{\frac{1}{a^2}-e^{-2\Phi}}= %\pm
\frac{2}{a\sqrt{ P(\r,r)}}\Big[
\sqrt{1-a^2} - \k \sqrt{1+a^2\k^2}\, \rho r \, 
\Big]\ ,  \la{9911}  \\
A&=\frac{2}{\sqrt{ P(\r,r)}}\Big[\sqrt{1+a^2\k^2}(c_1\rho dt+c_2 r d\vph) + \k \sqrt{1-a^2}(c_1r dt-c_2\rho d\vph)\Big]\ ,   \lab{299} \\
b&\equiv \frac{1}{\k}\sqrt{(1-a^2)(1+a^2\k^2)},\qquad \ \ \ \  c_1^2+c_2^2=1\ .\no 
\end{align}
This  solution  depends on  the parameter $a$ and also on a trivial parameter $c_1$   reflecting again  the freedom of  $U(1)$ 
electromagnetic duality    which  is  the symmetry of the equations following from \rf{4daction}
(we may always assume that $c_1 = c_2 ={1\ov \sqrt 2}$   without loss of generality). 
%The most symmetric choice is $c_1 = c_2 ={1\ov \sqrt 2}$ 
%when the undeformed \adst solution  has  $F_2=dA= \sqrt 2 ( d\rho\wedge dt+ dr\wedge d\vph)$, 
%i.e.  is  an electro-magnetic  Bertotti-Robinson solution. 

Let  us note that the   solution   for the scalar $C$  is of course defined up to a constant. Using this  
  the special   solution   corresponding to  $a=0$  can be   written  as\foot{Note that  
the infinite shift of the RR scalar effectively makes $C^2+4e^{-2\Phi}$ a function of the 
coordinates.  Even for $a\ne 0$ one can perform a constant shift to have $C=0$ for $\k=0$, 
making $C^2+4e^{-2\Phi}$ somewhat complicated. For $a=0$ the shift is required.}
%This solution depends on two free parameters, $a$ and $c_1$. Although the combination (\ref{29}) always 
%remains constant, this constant becomes singular for $a=0$, but the singularity can be eliminated by shifting $C$. 
%Thus, the case $a=0$ should be analyzed separately and the solution is
\ba
%\label{211}
%&ds^2=\frac{1}{1-\kappa^2\rho^2} \Big[-(1+\rho^2)dt^2+\frac{d\rho^2}{1+\rho^2}\Big]+ \frac{1}{1+\kappa^2r^2} \Big[(1-r^2)d\vph^2+\frac{dr^2}{1-r^2}\Big] \no
%\\
a=0: \ \ \ \ \ \ \ \ \ &e^{-2\Phi} =\frac{(1-\kappa^2\rho^2) (1 + \kappa^2 r^2)}{(1 - \kappa r \rho)^2 } \ , \ \ \ \ \ \qquad   C=0 \  , 
%\equiv \frac{(1-\kappa^2\rho^2) (1 + \kappa^2 r^2)}{G^2},\nonumber\\
\nonumber\\
&  A=\frac{2}{ 1 - \kappa r \rho }\Big[ c_1 (\rho  + \kappa r ) dt+c_2 ( r - \kappa \rho)  d\vph \Big] %\ ,
%\qquad c_1^2+c_2^2=1
\ . \label{1111}
\end{align}
Another special case  corresponds to $a=1$:
%Explicitly, for $a=1$ we find from \rf{299} (cf. \rf{111}) 
\ba %\label{299}
%ds^2&=\frac{1}{1 - \k^2\rho^2}\Big[-(1+\rho^2)dt^2+\frac{d\rho^2}{1+\rho^2}\Big]+\frac{1}{1 + \k^2r^2}\Big[(1-r^2)d\vph^2+\frac{dr^2}{1-r^2}\Big]
%\nonumber\\
a=1: \ \ \ \ e^{-2\Phi} &= %\frac{(1 - \k^2\rho^2) (1 + \k^2 r^2)}{ }\equiv
\frac{(1 - \k^2\rho^2) (1 + \k^2 r^2)}{  1 + \k^2\big( r^2-\rho^2 +r^2\rho^2\big)  }\ ,  %\ \ \ \ \  \ \ \      G^2(\r,r) \equiv 1 + \k^2\big( r^2-\rho^2 +r^2\rho^2\big)
\quad 
C=2\sqrt{1 -e^{-2\Phi}}= %\pm
-  \frac{2\k \sqrt{1+\k^2} }{\sqrt{ 1 + \k^2\big( r^2-\rho^2 +r^2\rho^2\big) } }  \, \rho\, r \ , \no\\
A&=\frac{2\sqrt{1+\k^2}}{\sqrt{1 + \k^2\big( r^2-\rho^2 +r^2\rho^2\big) } }\ \Big(c_1\rho dt+c_2 r d\vph\Big) \ . \lab{99}
\end{align}

%%%%%%%%%%%%%%%
\subsection{Symmetries  and limits of the solution}
\label{SubAdS2limits}

The free parameter $a$   should be fixed in order to  establish a relation to the  supercoset model
which depends just on $\k$.
 To understand  possible dependence of $a$ on $\k$    let us now 
discuss some properties and limits of the solution in \rf{199}--(\ref{299}). 
It turns out that it is invariant under certain sequences 
of dualities and analytic continuations:    %  Specifically,  let  consider two sets of transformations:
\begin{enumerate}[A.]
\item{\bf T--dualities}
\begin{enumerate}[1.]
\item{Perform T-dualities  along $t$ and $\vph$ directions.}
\item{Analytically continue the new    coordinates  
$(t,\vph)\rightarrow i(t,\vph)$ and  rescale
$(\rho,r)\rightarrow \ell \,(\rho,r)$, \  $\ell\equiv {\kappa}^{-1}$.}
\item{Replace the 2-form  potential,  appearing after the  T-dualities,   by an axion   via  tensor-scalar  duality in 4d.}
\item{Rescale the dilaton, the axion, and the Maxwell field to make $e^{-2\Phi}=1$ when $r=\rho=0$.}
\end{enumerate}
Then the  resulting geometry coincides with \rf{199}--(\ref{299}) upon the  identification
\bea\label{212}
a\rightarrow i\k a,\qquad \qquad \ell\rightarrow \k
\eea
\item{\bf Inversion of coordinates}
\label{ItemInvert4}
\begin{enumerate}[1.]
\item{Rewrite \rf{199}--(\ref{299}) in terms of $x\equiv 1/\rho$ and $y\equiv 1/r$.}
\item{Define $\ell\equiv {\kappa}^{-1}$ and ${\tilde L}=-i\ell L$   (we restore the overall scale $L$   in the metric  \rf{22}).}
\item{Rescale the dilaton, the axion, and the Maxwell field to make $e^{-2\Phi}=1$ when $r=\rho=0$.}
\end{enumerate}
The resulting geometry coincides with \rf{199}--(\ref{299}) upon the  identification (\ref{212}).
\end{enumerate}
\iffa 
%The demonstration of these %Detailed calculations 
%supporting these statements   statements 
 %showing that these transformations are indeed symmetries of the solution 
% are  given  in the appendix~\ref{AppSymm4D}. 
\fi

The transformation~B has  an  implication for the large $\kappa$ limit of \rf{199}--(\ref{299}):   if we want to send $\kappa$ 
to infinity while keeping the metric finite, then $L/\kappa$ must remain fixed. 
In the transformation~B this corresponds to sending 
$\ell=\k^{-1}$ to zero while keeping ${\tilde L}$ fixed; eq.~(\ref{212}) then implies that such limit leads to imaginary fluxes 
unless $a=0$. To see this more explicitly, let us  consider  the large $\kappa$ limit of \rf{199}--(\ref{299}) for $a\ne 0$:
\ba %\label{213}
ds^2&=\frac{L^2}{-\kappa^2\rho^2}
\Big[-(1+\rho^2)dt^2+\frac{d\rho^2}{1+\rho^2}\Big]+
\frac{L^2}{\kappa^2r^2}
\Big[(1-r^2)d\vph^2+\frac{dr^2}{1-r^2}\Big]
\nonumber\\
e^{-2\Phi} &=-\frac{\gamma^2 \kappa^2\rho^2 r^2}{a^2 (r^2-\rho^2)-2b r\rho+r^2\rho^2}\equiv
-\frac{\gamma^2 \kappa^2\rho^2 r^2}{P(\r, r) }, \qquad \qquad b=a\sqrt{1-a^2} \nonumber\\
C&=2\sqrt{\frac{\gamma^2}{a^2}-e^{-2\Phi}}=
\frac{\gamma}{\sqrt {P(\r, r)}  }\kappa\rho r,\nonumber\\
A&=\frac{2L\gamma }{\sqrt {P(\r, r)}  }\Big[a(c_1\rho dt+c_2 r d\vph)+\sqrt{1-a^2}(c_1r dt-c_2\rho d\vph)\Big]\ . \lab{2133}
\end{align}
Here we kept all coordinates fixed and rescaled the exponent of the 
dilaton and the RR fluxes by a free parameter~$\gamma$. It
% changed  $\alpha$ to $\gamma$ as tehre is an $\alpha$ in the SG action
is clear that no real value of this parameter makes $e^\Phi$   positive while keeping $C$ real. 
This argument breaks down only for $a=0$, when the expression for $C$ is  to be modified by an infinite constant shift. 
 For $a=0$ we  get 
\beaa\label{KinfA0mn}
ds^2&=&\frac{L^2}{-\kappa^2\rho^2}
\Big[-(1+\rho^2)dt^2+\frac{d\rho^2}{1+\rho^2}\Big]+
\frac{L^2}{\kappa^2r^2}
\big[(1-r^2)d\vph^2+\frac{dr^2}{1-r^2}\Big]
\nonumber\\
e^{-2\Phi} &=&-\gamma^2 \kappa^2,\qquad C=0,\qquad
A=\frac{2L\gamma }{(r\rho)^2}(c_1r dt-c_2\rho d\vph).
\eeaa
Setting $L=i\kappa$, $\gamma=1/L$, we find  AdS$_2\times$S$^2$   in the inverted coordinates.
%AAT
 This   suggests   that the parameter  $a$  should  vanish
 in this  large $\kappa$ limit.

%AAT
A different way of taking the large $\kappa$ limit of \rf{199}--(\ref{299}) is  found  by rescaling the coordinates  and $L$ as  follows 
(the  variables  with tildes  are to be kept fixed) 
\bea\label{FlatLimAdS2}
t=\frac{\tilde t}{L},\qquad \vph=\frac{\tilde\vph}{L},\qquad
\rho=\frac{\tilde\rho}{L},\qquad r=\frac{\tilde r}{L},\qquad \kappa=L{\tilde \kappa} \ , \  \ \ \ 
\k \to \infty, \ \  L \to \infty \ . 
\eea
Taking the limit $\k, L\rightarrow \infty$ in \rf{199}  we then   get  (omitting tildes) 
%v2
\ci{Arutyunov:2014cra}
\bea\label{KnownLimit}
ds^2=\frac{1}{1-\kappa^2\rho^2}
\big(-dt^2+d\rho^2\big)+
\frac{1}{1+\kappa^2r^2}
\big(d\vph^2+dr^2\big)
\eea
%AAT
The  T-dualities  in $t$ and $\varphi$  applied to  this metric  give dS$_2\times$H$_2$  space which  is  naturally a solution with a  constant 
dilaton. Then the simplest   choice   for 
  the dilaton that   represents a solution   together   with   the  metric (\ref{KnownLimit}) should   be
  %v2
   \ci{Arutyunov:2014cra}
\bea
e^{-2\Phi}=(1-\kappa^2\rho^2)(1+\kappa^2r^2) \ .  \la{223}
\eea
On the other hand, in the limit (\ref{FlatLimAdS2}) the dilaton in (\ref{999}) becomes
\bea
e^{-2\Phi}=\frac{(1-\kappa^2\rho^2)(1+\kappa^2r^2)}{1+(\kappa a)^2(r^2-\rho^2)-2a\sqrt{1-a^2}\kappa^2\, r\rho} \ . 
\eea
To match  \rf{223}   we  should thus  set $a=0$. 
This suggests that if $a$ is a function of $\k$ then one should have $a(\k\to \infty)\  \to 0$.
% $a$ is expected to be  a function of $\kappa$, 
%this

 %is consistent with the previous large $\kappa$ limit. 
%We conclude that $a$ must be equal to one the large $\kappa$ limit to keep al fluxes real. 

Another useful  limit  corresponds to  setting $\kappa= i$ in \rf{199}--(\ref{299})  (see \ci{HRT}).  Then  metric \rf{22},\rf{199}  
 becomes  flat, and the  dilaton \rf{999}  takes the form 
\bea
e^{-2\Phi}=\frac{(1+\rho^2)(1-r^2)}{(1+\rho^2)(1-r^2)-
(1-a^2)(\rho  + ir)^2} \ . 
\eea
%To keep this expression real, we must set $a=1$. 
This expression is real only if $a=1$   suggesting that one should  have   $a(\k\to  i)\  \to 1$. 
In this   case the dilaton becomes constant  as appropriate  
%AAT
for a  ``minimal"   choice  of the dilaton solution  in the case  a   flat  metric.  
There is also  a special way of  taking this  $\kappa= i$
 limit  by combining it with  a  rescaling of the coordinates
\bea\label{20}
t=\frac{x^+}{\eps}-\eps x^-,\qquad 
\vph=\frac{x^+}{\eps}+\eps x^-,\qquad \eps^2=\kappa^2+1\rightarrow 0 \ .
\eea
This leads to a  pp-wave  4d  metric  \ci{HRT}. In this limit the $F_2$  flux in   (\ref{299})   diverges  unless 
again $a=1$.

%%%%%%%%%%%%%%%%%%%%%%%%%%%%%%%%%%%
\iffalse 
{\bf Details for $a=1$ solution} (REMOVE LATER?)
The pp-wave limit rules out the solution with $a=0$ and selects   $a=1$  solution. We still have one problem and one curiosity:
\begin{enumerate}[(1)]
\item{$a=1$ solution  is that it does  not lead to right background in the limit $\k=\infty$ with $\k \rho, \k r$=fixed. }
\item{For  $a=1$ we have $C$   non-zero 
\bea \lab{c2d} 
C=\pm 
\frac{2k\sqrt{1 + \k^2}\  r\rho }{{[1 + \k^2( r^2-\rho^2 +r^2\rho^2)]^{1/2}}  }
\eea
and thus  there is a difference with $AdS_3$ case. 
But that may be suggesting that we need $C$   also in the $AdS_5$ case  as it is analogous 
to $AdS_2$ one   !?!?
}
\end{enumerate}
\fi 
%%%%%%%%%%%%%%%%%%%%%%%%%%%%%%%%%%%%%%%%%

%%%%%%%%%%%%%%%%%%%%%%%%%%%%%%%%%%%%%%%%%
\subsection{Choice  of $a(\k)$}
\label{SubAdS2a}

%In section \ref{SubAdS2soln} we found the most general solution (\ref{299}), which depends on two parameters $(a,c_1)$, and in section \ref{SubAdS2limits} we analyzed symmetries and limits of that geometry. Here we will use this information to propose the functional dependence $a(\kappa)$. 
%Let us now  propose the natural expression for $a$ as a function of $\k$. 
In the  previous subsection  we  discussed  the 
%AAT
natural 
   values of $a$   for  the  two  special  values of $\k$:
\begin{enumerate}[(i)]
\item{$\kappa=\infty$:   in this limit   the metric is  related (T-dual)  to an analytic continuation of  AdS$_2\times$S$^2$  and 
 the  %consistency of  the background %structure of $F_2$ implies  that  here 
 %implies that one should   
 simplest   choice is  to set  $a(\infty)=0$.}
\item{$\kappa= i$:  the 
 %AAT
assumption 
 that  $\Phi$  and $C$  should  remain real  within the family of solutions parametrized by $a$   implies   that $a(\k)$ should   satisfy 
$a( i)=1$.  }
\end{enumerate}
       %These two data points  suggest that $a$ must be a nontrivial function of $\kappa$,  
Let us  now propose a particular  function $a(\kappa)$ which  has  the required   limits
 and   is also   consistent   with  the   structure of the supercoset
 action.% (cf. \ci{DMV,ABF,HRT}).
    The deformed  supercoset action 
of \cite{DMV} depends  naturally on combination of  the    projectors 
$ \k  P_2   + \eta (P_1 - P_3)$  where 
%o on $\k$ and $\eta$ 
\bea \lab{1}
\eta = { \sqrt{ \k^2 + 1} -1 \over \k} \ , % \ \ \ \ \ \ \ \  \ \ \k= { 2 \eta \over \sqrt{ \eta^2 -1}} \ , 
\eea
and $P_k$ are projectors on the supergroup elements with $i^k$ charge under ${\rm  Z}_4$ transformations. 
The  string  sigma model  action and thus the background fields   should   then contain the two  parameters $\k$ and $\eta$ 
entering simply as a ratio. 
We conjecture that the solution \rf{199}--(\ref{299}) with  
$a(\kappa)$ given by 
\be\lab{2}
a(\k)= {\eta \over \k} = { \sqrt{ \k^2 + 1} -1 \over \k^2} = {1 \over \sqrt{ \k^2 + 1 } + 1} \ 
\ee
 should correspond to the   AdS$_2\times$S$^2$   supercoset model. 
Then 
\be \lab{3} a(0)={ 1 \over 2}\ ,\ \ \ \ \  \ \ \ \ \ a(i)=1 \ ,\ \ \ \  \ \ \ \ a(\infty)= 0\ ,   \ee
in agreement with the above discussion  of the two special limits. 
%We then match  the expected form of the background  in  both  (i)  $\k=i$  limit  and (ii)  
%$\k=\infty$  limit  with $\k \rho$ and $\k r$ fixed. 

 %%%%%%%%%%%%%%%%%%%%%%%%%%

\renewcommand{\theequation}{3.\arabic{equation}}
\setcounter{equation}{0}

\section{Deformation of AdS$_3\times$S$^3$}
\label{Sec6D}

 In the previous section we constructed a  supergravity solution   that  should  represent the background  underlying 
the $\kappa$-deformed AdS$_2\times$S$^2$ supercoset model. 
 The important ingredient  was the existence 
of a one-parameter family of solutions  with  the free parameter  $a$   which was then 
 fixed to  be  a specific function  of  $\kappa$  to match the corresponding limits of the  supercoset  construction. 

In this section we shall attempt to follow the same strategy for the $\kappa$-deformation of AdS$_3 \times $S$^3\times $T$^4$
space supported by RR 3-form flux. Compactifying on 4-torus we shall  use the  effective  6d action 
containing the dilaton $\Phi$, RR scalar  $C$ and RR 3-form field strenght $F_3$.  Starting with the $\k$-deformed \adstr   metric \ci{HRT} we   will 
find  again  a one-parameter family of solutions of  the three scalar equations of  the 6d theory. 
 It will turn out, however, 
   that only  two members of this family -- the analogs of the $a=0$ and $a=1$ solutions in  \rf{1111} and \rf{99} -- 
 can be  extended to solutions of  the full set of 6d equations  if one assumes that in addition to $\Phi$ and $C$ 
 only  one RR 3-form  is non-zero. 
% It is   likely 
%It is currently unclear if  one of them corresponds to the  supercoset model, 
%or if there should still exist a more general solution 

It is likely that there should  exist  a more general  solution (with an additional $F_5$ field in 10d or  an extra $F_3$ field in 6d) 
parametrized by   an arbitrary $a$ that  should  match the supercoset model for a special choice of $a=a(\k)$. 
%Such solution  appears  to  require   at least two independent RR 3-form  field strengths to be   non-zero. 

%%%%%%%%%%%%%%%%%%%%%%%%%%%%%%%
\subsection{One-parameter family of solutions of the scalar equations }

We shall   start with the following ``minimal" 6d  Lagrangian  representing a  reduction and truncation of type IIB  10d 
supergravity on 4-torus. As discussed in appendix \ref{AppEOM}, consistent truncation leads to the Lagragian
\bea
\label{6dL}
{\cal L}_6=e^{-2\Phi}\big[R+4(\nabla\Phi)^2\big]-
\frac{1}{12}F_{mnp}F^{mnp}-\frac{1}{2}(\del  C)^2 \ , 
\eea
supplemented by an additional constraint (\ref{Constr}) (see Appendix A):
\bea\label{ConstrMain}
\frac{1}{12}F_{mnp}F^{mnp}+\frac{1}{2}(\d C)^2=0 \ . 
\eea
The equations of motion for (\ref{6dL}) are also given  in appendix \ref{AppEOM}. 
The simplest   solution  is  \adstr supported by  the self-dual $F_{mnk}$   (with $\Phi$ and $C$ being trivial). 
We will be interested  in finding a solution 
for which the metric is  given by  the  $\k$-deformed  \adstr  metric  implied by the 
supercoset construction  \ci{ABF,HRT} %\foot{$L$ is again  the curvature scale that we set to 1 in what follows.} 
\be\label{31}
%L^{-2} 
ds^2= \frac{1}{1 - \k^2\rho^2}
\Big[-(1+\rho^2)dt^2+\frac{d\rho^2}{1+\rho^2}\Big]+
\rho^2 d\chi^2
\ \ + \ \ 
\frac{1}{1 + \k^2r^2}
\Big[(1-r^2)d\vph^2+\frac{dr^2}{1-r^2}\Big]+ r^2d\psi^2\ . \ \ 
\ee
As in the previous section we shall first focus on the three scalar equations: the trace of the Einstein's equation,  the RR scalar  one 
 and    the dilaton one   that can be organized as (cf. \rf{cc} and \rf{23}) 
\be \la{y}
R + 2\nabla^2\Phi   +  \frac{1}{2} e^{2 \Phi} \d_m C\d^m C  =0 \ ,\ \ \ \qquad  \nabla^2 C=0\ , \ \ \   \qquad 
\ \ \ \nabla^2\big( \frac{1}{2}C^2+e^{-2\Phi}\big)=0 \ . 
\ee 
We  begin by first solving them perturbatively in the small $\k$ limit  with $\k \r, \ \k^{-1} t $  being fixed as in \rf{24},\rf{25} 
when the metric becomes 
\bea\label{AdS3limit}
{ds^2}=\frac{1}{1-\k^2\rho^2}
\Big[-\rho^2dt^2+\frac{d\rho^2}{\rho^2}\Big]+
\rho^2 d\chi^2\ \ 
+\ \ (1-r^2)d\vph^2+\frac{dr^2}{1-r^2}+r^2d\psi^2 \ .
\eea
Expanding  in powers of $ \k\rho $ as in ~(\ref{PertExpDil})   we 
 find a unique solution (regular at $\kappa\rho=0$) which depends on one parameter $a$:
\ba
e^{-2\Phi}&=\frac{1-\k^2\rho^2}{P_2(\rho,r)}\ , \qquad   P_2(\r,r)  \equiv 
\big[1 - \k^2 (\rho r)^2\big]^2  +  2a^2  (2 r^2-1) (\k\rho)^2
- a^2  (2 r^2 - a^2) (\k \rho)^4  \ ,   \la{7777} \\
C&=
\frac{\sqrt{2}}{a\sqrt{(1-a^2)P_2(\rho,r)}}
\Big[1-2a^2+ (\k\rho)^2 (r^2 -a^2)\Big]
\ . \label{37}
\end{align}
This  small $\k$  solution can be generalized to the  arbitrary  $\k$ solution of the three { scalar equations} \eqref{y}:
\ba
%\label{366}
e^{-2\Phi}&=\frac{(1-\k^2\rho^2)(1+\k^2r^2)}{P_2(\rho,r)}\ ,\la{3666}
\\
P_2(\r, r)&\equiv \big[1- \k^2 (\rho r)^2\big]^2 +2a^2\k^2 \big[r^2-\rho^2+2 (\rho r)^2\big] 
+
2\k^4a^2(\rho r)^2(r^2-\rho^2-2) +{\k^4a^4}(\rho^2+r^2)^2\ ,  \no \\
\label{366}
C&=\sqrt{\frac{1}{2a^2(1-a^2)}- 2e^{-2\Phi}}=
\frac{\sqrt{2}}{a\sqrt{(1-a^2)P_2(\r,r)}}
\Big[1-\k^2 (\rho r)^2-a^2(2-\k^2\rho^2+\k^2 r^2)\Big] \ . 
\end{align}
%This solution is unique and depends on one free parameter $a$. 
The same choice \rf{2} for $a(\k)$ as in the  \adst case  then  gives  us a solution which is consistent with both 
$\k=\infty$ and $\k=i$ limits.

It is interesting to note  a  relation between  the quadratic polynomial $P$ in the deformed  \adst   solution \rf{199}--(\ref{299})
and  the quartic polynomial $P_2$ in \rf{366}.  If we define the analog of $P\equiv P_- $ in (\ref{999})
with\footnote{Changing the sign of $b$ maps  \rf{199}--(\ref{299}) into another solution 
 provided one also   changes  the relative sign of the two terms in the 
1-form field in \rf{299}.}
 $b \to - b$  as $P_+$ then we   observe that 
$P_2$ can be written as a product of $P_+$ and $P_-$, i.e. 
\bea
P_2 = P_+ P_- \ , \qquad  P= P_- \ , \ \ \ \quad 
P_\pm\equiv 1 + \k^2a^2  (r^2-\rho^2) \pm 2 \k\sqrt{(1-a^2)(1+a^2 \k^2)} \ r\rho + \k^2 r^2\rho^2 
 \  .    \ \  \la{38} 
\eea

Attempting to extend this one-parameter solution  of  the scalar equations to a solution of the 
full set of 6d  equations following from \rf{6dL}--\rf{ConstrMain}  using  an ansatz-based approach suggests that this is 
possible only for the special values  $0$ and $1$  of the  parameter $a$. 
We shall also see another indication of this obstruction 
from the algebraic constraints on the 3-form stress tensor  discussed   in the next subsection. 
 
 %Unfortunately however, in the present case solutions with $a\ne\{0,1\}$ of the scalar 
%equations cannot be extended to solutions of \eqref{6dL}. 

% To understand this and simplify the analysis for general $\kappa$ we first explore 
%when a certain six-dimensional configuration of the metric and other fields can be sourced by a single 
%three-form field strength.

%%%%%%%%%%%%%%%%%%%%%%%%%%%%%%%%
\subsection{Existence of  %a middle-form
a  field strength for  a given stress  tensor:
  Rainich conditions % in higher dimensions
  \label{6dRainich}}

The   question we are facing is  how to  find a 3-form flux supporting  (together  with the scalar fields) a given 
metric through the Einstein equation, i.e. how  to find a solution  for the flux  given a  specific form of  its   stress tensor. 
In general,  the  question  is when  some field configuration (i.e. some metric as well as other fields) 
can be sourced by an   $n$-form field in $d=2n$ dimensions.

In the four dimensional Einstein-Maxwell  theory  this question  was   addressed  long 
ago \cite{Rainich_original,Misner:1957mt, Torre:2012hw, Torre:2013nia}:
in order for some   stress tensor $T_{mn}$  implied by the Einstein's equations 
to   be generated by a  Maxwell field  strength  $T_{mn}$   should be traceless   and also 
its third power should be traceless as well (a brief derivation   of this fact  is given in appendix  \ref{AppRainich}).
% we include a brief derivation of this fact.

Here we  find  the analogous conditions in six  dimensions 
(the generalization to higher dimensions  is also  straightforward). 
Let us  consider the stress tensor 
of a 3-form field strength
\bea\label{6DStressFF}
{T_m}^n=F_{mkl}F^{kln}- \frac{1}{6}\delta_m^n F_{skl}F^{kls} \ .
\eea
Direct calculation shows that  it satisfies 
\bea\label{OddTrace0}
\mbox{tr}\, T=0\ ,\qquad \mbox{tr}\, T^3=0\ ,  \qquad \mbox{tr}\, T^5=0 \ .
\eea
Thus   given  a six-dimensional   background (metric, dilaton, etc.)  and computing 
the effective    stress tensor $T_{mn}$ in the right-hand side of the Einstein equation   that  should 
be representing the contribution of the 3-form field, 
 this  $T_{mn}$ should satisfy eq.~\rf{OddTrace0}
in order  for $F_{mnk}$ to exist. This is a necessary condition, which in general may not  be a sufficient one.

 \iffa 
  described  by a metric, dilaton and RR  scalar and vector fields may be supported by a 
three-form field strength if the effective stress tensor defined by Einstein's equation 
\bea
R_{\mu\nu} -\frac{1}{2}g_{\mu\nu} R = T^\text{scalars}_{\mu\nu}+T^\text{1-form}_{\mu\nu}+T^\text{2-form}_{\mu\nu} 
\longrightarrow 
T^\text{2-form}_{\mu\nu} =R_{\mu\nu} -\frac{1}{2}g_{\mu\nu} R - T^\text{scalars}_{\mu\nu}-T^\text{1-forms}_{\mu\nu} \ ,
\label{effectiveT}
\eea
satisfies these conditions \eqref{OddTrace0}. We note here that while these conditions are necessary, it is by no 
means clear that they are sufficient. 

It is not difficult to  find an algebraic equation  for $T_{mn}$  which is  subject to the conditions \eqref{OddTrace0}. 
Any $6\times 6$ matrix obeys  sixth order characteristic equation
\bea\la{311} 
T^6+\sum_{k=0}^5 c_k T^k=0,
\eea
and  eqs.~\eqref{OddTrace0} imply that upon diagonalization  $T_{m}^n$ should  take the form 
\bea
T=\mbox{diag}(\l_1,-\l_1,\l_2,-\l_2,\l_3,-\l_3) \ .
\eea
Then  the coefficients $c_k$  in \rf{311}   can be found  explicitly and one ends with the following 
equation for the  corresponding stress tensor $T$ 
\bea
T^6-\frac{1}{2}(\mbox{tr}\,T^2)\,T^4-
\frac{1}{8}
\left[2\,\mbox{tr}\,T^4-(\mbox{tr}\,T^2)^2\right]T^2
-\frac{1}{6}\mbox{tr}\, T^6+
\frac{1}{8}\mbox{tr}\, T^4~\mbox{tr}\, T^2-
\frac{1}{48}(\mbox{tr}\, T^2)^3=0 \ .
\eea
As one can see  on explicit examples of three-form field strength $F_{mnp}$ in (\ref{6DStressFF}), 
 in general $T$ does not  satisfy any equation of degree lower than six.
\fi

%Using the same effective stress tensor \eqref{effectiveT} it is possible to ascertain whether the three-form field strength (and, in general, 
%the middle-dimensional field strength) has additional special properties. 
Some  additional constraints may appear for special  choices of the   field strength. %  that produces given  $T_{mn}$. 
For example, for an (imaginary)-self-dual field strength 
we find that
\beaa
4d:\ \ \ ~{T_m}^n=0\ , \qquad 
~~\qquad~~
6d:\ \ \ ~T^2=\frac{1}{6}\mbox{tr}\, T^2 \ .
\eeaa
A similiar analysis  implies  that the 
necessary conditions that some 10d   symmetric 2nd rank tensor may be the  stress tensor  of 
a five-form field strength are
\bea
\label{10dRainich}
\mbox{tr}\, T=0,\qquad \mbox{tr}\, T^3=0, 
\qquad \mbox{tr}\, T^5=0,\qquad \mbox{tr}\, T^7=0,
\qquad \mbox{tr}\, T^9=0 \ .
\eea
These relations hold, in particular,  for AdS$_5\times M^5$ solutions, where $M^5$ is an Einstein space.

%%%%%%%%%%%%%%%%%%%%%%%%%%%%%%%%%%%%
\subsection{Complete solutions}

Starting with the metric  ~\eqref{AdS3limit}  and    the dilaton  \rf{7777} and RR scalar 
  \eqref{37}  one can find explicitly the expected    stress tensor for the 3-form RR field  $F_3= d C_2$
\beaa\label{TmnDef}
T_m{}^n\equiv  e^{-2\Phi}(R_{m}{}^n+2\nabla_m\nabla^n\Phi) -
\frac{1}{2}\big[\d_m C\d^n C-\frac{1}{2}\delta_m^n(\nabla C)^2\big] = 
\frac{1}{4}\big({F}_{mpq}{F}^{npq}-
\frac{1}{6}\delta_m^n{F}_{spq}{F}^{spq}\big)
\ .\ \ \ 
\eeaa
Direct calculation shows that it satisfies the non-trivial (last two) relations  in \eqref{OddTrace0} 
only for  the special values $a=0,1$
of the parameter in   \eqref{37}. 
Thus (\ref{37}) 
can be supported by the 3-form  flux only in these two special  cases.
% in agreement with what was  suggested by a  direct ansatz-based analysis.
It may be possible  to go around   this   problem  by allowing for two  non-vanishing independent 3-form fields 
in the  reduced 6d Lagrangian \rf{6dL}. We will not attempt to study this option here. 

%It appears however that solutions exist if one further 
%extends the six-dimensional Lagrangian by another three-form field strength.

 The corresponding  small $\k$ limit solutions (supplementing the metric \eqref{AdS3limit})   are:
\ba
%\underline{a=0}:&&\nonumber\\
a=0: \ \ \ \ \ e^{-2\Phi}&=\frac{1-\k^2\rho^2}{[1-(\k\rho r)^2]^2},\qquad \qquad \qquad   \ C=0,  \label{666} \\
C_2&=\frac{\rho^2}{1-(\k\rho r)^2}\big[dt+\k(1-r^2)d\vph\big]\we 
\big[d\chi+\k r^2d\psi\big] -r^2d\vph \we d\psi\ , \no \\
%
%\underline{a=1}:&&\nonumber\\
a=1: \ \ \ \ \  e^{-2\Phi}&=\frac{(1-\k^2\rho^2)(1+\k^2r^2)}{[1-(\k\rho)^2(1- r^2)]^2},\qquad \ \ \  C=0,   \label{777} \\
C_2&=\frac{1}{1-(\k\rho)^2(1- r^2)}\Big[\rho^2 dt\we d\chi-
\k\rho^2(1-r^2)dt\we d\vph +\k(\rho r)^2 d\chi\we d\psi-r^2 d\vph \we d\psi\Big]\  .\no 
\end{align}
%This bifurcation seems reasonable since in the limit (\ref{AdS3limit}) there is a symmetry under interchange of $\vph$ and $\psi$ that swaps the two expressions.
These small-$\k$ limit solutions can be extended to
 solutions  with  general $\k$ by making ans\"atze that dress the solutions \eqref{666} and \eqref{777}
with numerator and denominator functions of $\kappa$, $r$ and $\rho$ which become unity in the small $\kappa$ limit.
  The resulting exact solutions are found to be (cf. \rf{1111},\rf{99})
\ba
%\underline
&{a=0}: \ \ \  \ \ \ 
e^{-2\Phi}=\frac{(1-\k^2\rho^2)(1+\k^2r^2)}{[1-(\k\rho r)^2]^2},\ \ \ \qquad\qquad \  \  \ C=0,   \label{316}  \\
%\label{Full3_a0}
%\label{aEq0Flux}
%e^{-2\Phi}&=&\frac{(1 - \k^2\rho^2)(1 + \k^2r^2)}{[1- ( \k \rho r)^2]^2},\qquad C=0\nonumber\\
&C_2=
%\frac{1}{1-(\k\rho r)^2}\Big[\rho^2(1+\k^2r^2) dtd\chi+
%\k r^2(1+\rho^2)dtd\psi \\
%& +\k\rho^2(1-r^2)d\vph d\chi- r^2(1-\k^2\rho^2)d\vph d\psi\Big]\\
\frac{1 }{1-(\k\rho r)^2} \Big[     \r^2  \big(dt+\k d\vph\big)  \we  \big(d\chi+\k r^2d\psi\big) -
r^2  \big(d\vph-\k dt\big)\we  \big(d\psi+\k\rho^2d\chi\big)\Big]  \no  \\
%\lab{a13d}
%\underline
&{a=1}: \ \ \ \ \ \  e^{-2\Phi}= \frac{(1-\k^2\rho^2)(1+\k^2r^2)}{\big[ 1+\k^2 (r^2 -\rho^2 + r^2 \rho^2)\big]^2},\qquad
%P(\rho,r)=\big[ 1+\k^2 (r^2 -\rho^2 + r^2 \rho^2)\big]^2 \ , 
   \qquad C=0,\la{72}  \\ 
%\label{Full3_a1}
%   {ds^2}&=&\frac{1}{1 - \k^2\rho^2} \Big[-(1+\rho^2)dt^2+\frac{d\rho^2}{1+\rho^2}\Big]+ \rho^2 d\chi^2
%\nonumber\\
%&&
%+ \frac{1}{1 + \k^2r^2} \Big[(1-r^2)d\vph^2+\frac{dr^2}{1-r^2}\Big]+r^2d\psi^2\nonumber
%
%
%
&C_2=\frac{\sqrt{1+\k^2}}{  1+\k^2 (r^2 -\rho^2 + r^2 \rho^2)}\Big(\rho^2 dt\we  d\chi + \k\big[r^2-\rho^2+ (\rho r)^2\big]dt\we 
d\vph +\k(\rho r)^2 d\chi \we d\psi - r^2 d\vph\we  d\psi\Big) \ .  \no 
%\label{72}
\end{align}
There are also solutions with flipped signs of $t,\chi,\vph,\psi$. % ($2^4$ combinations).

Let us note that, as in the 4d  case \rf{3322},   the undeformed \adstr metric (i.e. \rf{31} with $\k=0$) 
can be supported by a   one-parameter family of  2-form potentials
 %however, unlike the $\k=0$ case where the 
%two-form potential can be continuously shifted between the AdS$_3$ and $S_3$ factors, 
\bea
\label{c1c2family}
C_2=\sqrt 2 \big( c_1\rho^2 dt\we  d\chi+c_2r^2 d\vph \we  d\psi\big)\ ,\qquad\qquad 
c_1^2+c_2^2=1 \ .
\eea
However, this freedom does not extend   to   the case of 
$\k\ne 0$  with nontrivial $\Phi$ and $C$. 
This is related to a different structure of the   ``electro-magnetic" duality group  that acts on the  3-form field strength:
 in 4d this is $SO(2)$ that rotates $(c_1,c_2)$ and in 6d   this is  $Z_2$.

%%%%%%%%%%%%%%%%%%%%%%%%%%%%%%%%%%%%%

\subsection{Symmetries and limits of the solution} % and selection of $a$}
%%%%%%%%%%%%%%%%%%%%%%%%%%%%

%Solutions described in the last subsection have some interesting symmetry properties, which may be useful in identifying $a(\k)$ even without explicitly 
%identifying the three-form flux(es) supporting the solution.
% 
Let us  now discuss   some properties of the solutions \rf{3666},\rf{366} and \rf{316},\rf{72} corresponding to the metric \rf{31}. 
\begin{enumerate}[A.]
\item{\bf Swap of the coordinates on the sphere}\\
The metric (\ref{31}) is invariant under swapping of the angles on S$^3$ together with a redefinition of $r$: 
\bea\label{AngleSwap}
\psi\leftrightarrow\vph,\qquad \qquad 
r\rightarrow \sqrt{\frac{1-r^2}{1+\k^2 r^2}}
\eea
One can check  that the scalar fields in  \rf{3666},(\ref{366}) remain invariant  provided one also  transforms  $a$  as  
\bea
a\ \rightarrow\ \frac{1-a}{1+a\k^2} \ . 
\eea
In particular, the points $a=0$ and $a=1$ are interchanged, and, in fact, 
the complete $a=0$ solution \rf{316}  is interchanged with  the $a=1$   solution \rf{72}. 
\item{\bf T--dualities}\\
As in the \adst case, we can perform a sequence of  transformations:
\begin{enumerate}[1.]
\item{T--dualize  along $t$ and $\vph$ directions.}
\item{Continue the new  coordinates as 
$(t,\vph)\rightarrow i(t,\vph)$ and  rescale
$(\rho,r)\rightarrow \ell (\rho,r)$, \ $\ell\equiv {\k}^{-1}$.}
\end{enumerate}
This sequence  maps 
%Application of this procedure maps 
the $a=0$ solution \rf{316}  back to itself (after an appropriate rescaling of coordinates). The $a=1$  background \rf{72}  
is mapped into a solution with 
imaginary fluxes, which cannot be made real by further analytic continuations.
 %The detailed analysis is included in 
%appendix~\ref{AppSymm6D}. {\bf DO WE NEED THIS ?!} 
%
\item{\bf Inversion of coordinates}\\
As in the \adst case, the limit  $\k=\infty$ simplifies 
after a sequence of duality transformations and analytic continuations:
\begin{enumerate}[1.]
\item{Rewrite \rf{199}--(\ref{299}) in terms of $x\equiv 1/\rho$ and $y\equiv 1/r$.}
\item{T--dualize along $\psi$ and $\chi$.}
\item{Define  ${\tilde L}=-i\ell L$,  \ $\ell\equiv {\k}^{-1}$. }
\end{enumerate}
%The details of this procedure are presented in the appendix \ref{AppSymm6Dinv}, where 
One can  show that  then  %     rit is shown that the 
the RR fields  become complex unless $a=0$. 
%MAYBE SAY MORE ABOUT THE TRANSFORMATIONS OF THE DILATON (not sure it is necessary --R).
\end{enumerate}

%AAT
Thus as in the \adst   case  the 
 large-$\k$ limit  appears to prefer the $a=0$ solution.
% However, as in the \adst case,  we cannot set $a(\k)=0$ for all $\k$ since this   will be in an  apparent 
At the same time,   the $\k=i$ or pp-wave limit  \cite{HRT}  appears to prefer  the $a=1$ solution. 
  Namely,   if we  consider again  the limit \rf{20} 
%\bea\label{78} t=\frac{x^+}{\eps}-\eps x^-,\qquad 
%\vph=\frac{x^+}{\eps}+\eps x^-,\qquad \eps^2=\k^2+1\rightarrow 0,
%\eea
%and set $\k=i$, 
then  $C_2$ in (\ref{316}) diverges, while  $C_2$  in  (\ref{72}) remains finite and real, i.e. 
\be
a=1: \ \ \ \  \ \ \ \ C_2= {1 \over (1+ \rho^2)(1-r^2)  }  \big(  \rho^2 d \chi + r^2 d \psi \big) \we d x^+  \ .
\ee
Comparing this  with eqs.~(3.28)~and~(3.29)  in  \cite{HRT} giving the two-form potential in this limit
%\be  C_2 = \big( \sin^2 \alpha \ \cosh^2 \beta  \ d \psi + \cos^2 \alpha \ \sinh^2 \beta \  d \psi\big) \we d x^+  \ ,\ee
and accounting for the coordinate change %($\rho=\tan \alpha, \ r=\tanh \beta, \ \ \psi\to \chi$) 
we find a perfect match.

The value  $a=1$   for $\k= i $ is also singled out by comparing
 the  corresponding  limits of the dilatons in \rf{316} and \rf{72}   % and (\ref{Jul29a1})
\bea\label{kIdila01}
\left.e^{-2\Phi}\right|_{a=0}=
\frac{(1+\rho^2)(1-r^2)}{[1+(\rho r)^2]^2},\qquad\qquad \quad 
\left.e^{-2\Phi}\right|_{a=1}=
\frac{(1+\rho^2)(1-r^2)}{(1+\rho^2)^2(1-r^2)^2}=
\frac{1}{(1+\rho^2)(1-r^2)}
\eea
with the expression for the  
%AAT
natural value of the dilaton found directly in this limit in \cite{HRT} (see eqs.~(3.16), (3.22) and~(3.26) there). 
%\bea e^{-2\Phi}=(1-\sin^2\alpha)(1+\sinh^2\beta)=\frac{1}{(1+\tan^2\alpha)(1-\tanh^2\beta)}=\frac{1}{(1+\rho^2)(1-r^2)} \ .
%\eea
%(see also eqs. (3.16) and (3.22) in \cite{HRT}). 

We  conclude that, as in the \adst case, the 
limits $\k= \infty$ and $\k=i$  appear to  select   two  different values of $a$,
 suggesting  that  there should  exist an interpolating solution 
with  $a=a(\kappa)$. 
While the six-dimensional Rainich conditions discussed in sec.~\ref{6dRainich}
rule out such a solution supported by a single 3-form flux, a preliminary  investigation suggests that  there may  exist  a  6d  supergravity 
solution  with  two different 3-form fields being non-zero.

%%%%%%%%%%%%%%%%%%%%%%%%%%%%%%%%%%%%
\subsection{A generalization: 2-parameter deformation}
%AT

It was shown in \cite{HRT} that the $\kappa$-deformation of the AdS$_3\times $S$^3$  metric 
 corresponds  to  a special case of the general 2-parameter Fateev 
model \cite{fat} which is also  the same as the 2-parameter family of classically integrable  bi-Yang-Baxter
 sigma models constructed 
in \cite{k2, k3}. The  corresponding   deformed \adstr  metric can be  written 
as
\ba
\label{325}
&ds^2=\frac{1}{F(\r)}
\Big[-(1+\rho^2)\big[1+\k_-^2(1+\rho^2)\big]dt^2+ \frac{d\rho^2}{1+\rho^2} + 
\rho^2 (1-\k_+^2\rho^2)d\chi^2+2\k_-\k_+\rho^2(1+\rho^2)dtd\chi\Big]
\nonumber\\
&\qquad +
\frac{1}{\tilde F(r)}
\Big[ \ \ (1-r^2)\big[1+\k_-^2(1-r^2)\big ]d\vph^2  + \frac{dr^2}{1-r^2} +r^2(1+\k_+^2r^2)d\psi^2+2\k_+\k_-r^2(1-r^2)d\psi d\vph\Big]\ , 
\nonumber\\
&\qquad \qquad F=1+\k_-^2(1+\rho^2) - \k^2_+\rho^2,\qquad\qquad 
{\tilde F}=1+\k_-^2(1-r^2)  + \k_+^2r^2 \ . 
\end{align}
For  $\k_-=0, \ \k_+ =\k$ we get  back to the metric \rf{31}. 
There is  no $B$-field. The   supercoset model  with this bosonic part was constructed in \cite{ben}. 
%A supercoset construction for this two-parameter deformation was identified in 
Similarly to the  case of the $\k$-deformed  AdS$_3\times $S$^3$
metric, it should  thus be possible to extend the metric \rf{325}  to a full supergravity solution.
% of the  six-dimensional supergravity. 

Indeed, we  found  the  following 
 generalizations of the $a=0$   \rf{316} and $a=1$  \rf{72}  solutions  with both $\k_+$ and $\k_-$: 
\ba
&a=0: \ \ \ \ 
e^{-2\Phi}= \frac{F(\r) {\tilde F(r)}}{[P(\r,r)]^2} \ , \ \ \ \ \ \   P \equiv  1+\k_-^2-(\k_+^2-\k_-^2) r^2\rho^2
%\frac{F{\tilde F}}{[1+\k_-^2+(\k_+^2-\k_-^2)(r^2-\rho^2+r^2\rho^2)]^2}\equiv \frac{F{\tilde F}}{P^2}
\ ,\qquad \ \ \ \ \qquad \qquad C=0\ , \no %\la{22630}
\\
&C_2=\frac{\sqrt{1+\k_-^2}}{P(\r,r)}\Big[(1+\rho^2) dt\we d\chi+(1-r^2) d\vph\we  d\psi  +
\k_+(1+\rho^2)r^2 dt\we d\psi -\k_+\rho^2 (1-r^2)d\chi\we  d\vph
\nonumber\\
&\Big. \qquad \qquad \qquad \quad
+\k_-(1+\rho^2)(1-r^2)dt\we d\vph-\frac{\k_-(1+\k_+^2)}{1+k_-^2}r^2 \rho^2 d\chi \we d\psi\Big]\ ,  \la{32260}
\\
& a=1: \ \ \ \ 
e^{-2\Phi}= \frac{F(\r) {\tilde F(r)}}{[P(\r,r)]^2} \ , \ \ \ \ \ \   P \equiv  1+\k_-^2+(\k_+^2-\k_-^2)(r^2-\rho^2+r^2\rho^2)
%\frac{F{\tilde F}}{[1+\k_-^2+(\k_+^2-\k_-^2)(r^2-\rho^2+r^2\rho^2)]^2}\equiv \frac{F{\tilde F}}{P^2}
\ ,\qquad \ \ C=0\ , \no %\la{2263}
\\
&C_2=\frac{\sqrt{1+\k_+^2}}{P(\r,r)}\Big[\rho^2 dt\we d\chi-r^2 d\vph\we  d\psi+
\k_-(1+\rho^2)r^2 dt\we d\psi -\k_-\rho^2 (1-r^2)d\chi\we  d\vph\Big.
\nonumber\\
&\Big. \qquad \qquad \qquad \quad 
+\frac{\k_+(1+\k_-^2)}{1+\k_+^2}(1-r^2)(1+\rho^2)dt\we d\vph-\k_+r^2\rho^2 d\chi \we d\psi\Big]\ . \la{3226}
\end{align}
As in the \adst and \adstr  cases discussed above, it is natural to expect that there should exist a one-parameter family of solutions including  \rf{32260} and \rf{3226}  as special cases. 

Solutions \rf{32260} and \rf{3226} are interchanged by the transformation
\bea\label{A01map327}
\rho\rightarrow i\sqrt{1+\rho^2},\quad r\rightarrow 
\sqrt{1-r^2},\quad t\leftrightarrow \chi,\quad
\vph\leftrightarrow \psi,\quad
\k_+\leftrightarrow \k_-\,.
\eea
The  invariance of the metric \rf{325} under the map \rf{A01map327}  was  noted  in \cite{ben}.

\renewcommand{\theequation}{4.\arabic{equation}}
\setcounter{equation}{0}
%%%%%%%%%%%%%%%%%%%%%%%%%%%%%%%%%%%%%%%%%%%%%%%%%%%%%%%%
%\newpage 
\section{Deformation of AdS$_5\times$S$^5$}

The extension of the $\k$-deformed \adss   metric and $B$-field  \ci{ABF}  to a full supergravity solution  turns out  to be  
more challenging than in  the above lower-dimensional  cases. This is  due, in particular, to  the lack of isometries, i.e.  
 a non-trivial dependence on the two extra  angular coordinates. 
While we   will not  find  a complete 
solution, in this section we shall discuss some of its features and  draw analogies   with the \adst and \adstr   cases. 
 %the lower-dimensional cases.

Assuming a particular structure  of  the  RR fluxes we shall find two different 
solutions to the   scalar equations which are the counterparts of 
 the $a=0$  \rf{316} and $a=1$ \rf{72}  solutions     in the AdS$_3\times$S$^3$ case
 (we shall thus refer to them as the ``$a=0$'' and ``$a=1$" solutions). 
%For this reason we shall refer to them as the $a=0$ and $a=1$ solutions. 
%
To construct them it will be useful to   switch  to a T-dual frame  where there is no $B$-field. 
We shall find that in  this  frame  both  solutions   have 
vanishing RR scalar, $C=0$.  However, 
in contrast to the  AdS$_3\times$S$^3$ case we have  been unable to 
find  a one-parameter  family connecting these two special      solutions.
Moreover, the  10d  algebraic Rainich  conditions discussed in sec.~\ref{6dRainich} imply that %, 
%in the T-dual frame, 
these solutions cannot be supported solely by a 5-form flux, i.e.  one should    excite 
other fluxes  as well. We leave the study of this possibility for the future.

%\subsection{Solutions  of  the scalar equations}
\def \vphi {\varphi}

Our starting point   will be   the deformed AdS$_5\times$S$^5$ metric and $B$-field  corresponding  \cite{ABF}
to the $\k$-deformed  supercoset model of \ci{DMV}\foot{Let us mention that 
 the detailed form  of the   model of \ci{DMV}  depends on a choice of the matrix $R$  and there  are several  possibilities  discussed  in 
\ci{Delduc:2014kha}  (in the \adstr case there are  two choices related   to $\k_-=0$ or $\k_+=0$  in \rf{325}, see \ci{ben}). 
Here we shall  consider only the original choice in \ci{DMV,ABF}.} 
\ba
ds^2&=f(\r) \Big[-(1+\rho^2)dt^2+\frac{d\rho^2}{1+\rho^2}\Big]+v(\r, \zeta)\,  
\rho^2(d\zeta^2+c_\zeta^2 d\psi_1^2)+\rho^2s_\zeta^2 d\psi_2^2\no \\
& \ \ +{\tilde f}(r) \Big[(1-r^2)d\vphi^2+\frac{dr^2}{1-r^2}\Big]+{\tilde v}(r, \theta)\,  r^2(d\theta^2+c_\theta^2 d\phi_1^2)+r^2s_\theta^2 d\phi_2^2   
\label{41} \ , \\
%
%%%%%%
\iffalse
%%%
B&=\frac{1}{2}\k \Big[  v(\r,\z) \,  \rho^4\sin 2\zeta d\psi_1\wedge d\zeta-
 {\tilde v}(r, \theta)\, r^4\sin 2\theta d\phi_1\wedge d\theta \Big] 
 %%%
 \fi
 %%%%%%
 B&=\frac{1}{2}\k \Big[  2 v(\r,\z) \,  \rho^4 s_\zeta c_\zeta d\psi_1\wedge d\zeta-
 2 {\tilde v}(r, \theta)\, r^4 s_\theta c_\theta d\phi_1\wedge d\theta \Big] 
 \ , \la{411}
\\  f&=\frac{1}{1-\k^2\rho^2},\quad
{\tilde f}=\frac{1}{1+\k^2r^2} \ , \qquad 
v=\frac{1}{1+\k^2\rho^4s_\zeta^2},\quad
{\tilde v}=\frac{1}{1+\k^2r^4s_\theta^2}
\nonumber \ ,
\end{align}
where we used the shorthand notation
$
s_x= \sin x \ , \ \ 
c_x= \cos x $. 

To put this  background  on the equal footing with the  above low-dimensional solutions, it is convenient to remove the 
NS-NS $B$-field by performing the  T-duality  along the two compact coordinates $\phi_1$ and $\psi_1$ 
  which  gives a non-diagonal metric 
%  gives $B=0$   while the metric acquires off-diagonal components:
\ba\label{42}
ds^2&=f\Big[-(1+\rho^2)dt^2+\frac{d\rho^2}{1+\rho^2}\Big]+v\rho^2 d\zeta^2+\rho^2s_\zeta^2 d\psi_2^2+{\tilde f}\Big[(1-r^2)d\vphi^2+\frac{dr^2}{1-r^2}\Big] \no \\
%%%%%%
\iffalse
%%%
&+{\tilde v}r^2 d\theta^2+r^2s_\theta^2 d\phi_2^2+\frac{1}{v\rho^2c_\zeta^2} \Big(d\psi_1+
\frac{1}{2}kv \rho^4\sin 2\zeta d\zeta\Big)^2+
\frac{1}{{\tilde v}r^2c_\theta^2} 
\Big(d\phi_1-\frac{1}{2}k{\tilde v}r^4\sin 2\theta
d\theta\Big)^2 %\\
%%%
\fi
%%%%%%
&+{\tilde v}r^2 d\theta^2+r^2s_\theta^2 d\phi_2^2+\frac{1}{v\rho^2c_\zeta^2} \Big(d\psi_1+
kv \rho^4 s_\zeta c_\zeta d\zeta\Big)^2+
\frac{1}{{\tilde v}r^2c_\theta^2} 
\Big(d\phi_1- k{\tilde v}r^4 s_\theta c_\theta d\theta\Big)^2  \ .%\\
%
%B&=&0.\nonumber
\end{align}
In the absence of the $B$-field the dilaton equation is 
\bea\label{44}
R+ 4 \nabla^2 \Phi - 4 (\partial \Phi)^2 =0.
\eea
%The other equations depend on the structure of the Ramond--Ramond fluxes and we shall {\it assume} that 
We shall  assume that   in addition to the metric and the dilaton  only the RR scalar 
$C$ and  the 5-form field $F_5=C_4$ are excited.
Then  for a given metric  the scalars 
 $C$ and ${\Phi}$   must satisfy an over-constrained system  of the three  equations -- 
 (\ref{44}) as well as  the  RR scalar equation and the
  trace of the Einstein  equation:
\bea
\nabla^2 C=0,\qquad \qquad R + 2 \nabla^2 \Phi + 2 e^{2\Phi}( \partial  C)^2 =0 \ .
\eea
One of these three   may be replaced with  (cf. \rf{cc}, \rf{23}, \rf{y})
\be \lab{eqss}
   \nabla^2  \Big( C^2 + e^{-2 \Phi}   \Big)  =0 \ .
\ee 
%which is just a combination of the three scalar equations (see (\ref{cc})).
Since the metric  \rf{42} was obtained from the deformed AdS$_5\times$S$^5$  NS-NS background by the application of T-dualities, 
it  should have  a nontrivial dilaton even in the absence of the deformation
(we shall   denote the dilaton in T-dual frame with tilde)
\bea \la{47}
e^{-2\td  \Phi}\Big|_{\k=0}=(\rho c_\zeta \, r c_\theta)^2 \ . 
\eea
In  the general case  we  may  then  parametrize  the dilaton as (cf. \rf{999},\rf{7777})
\bea \la{48}
e^{-2\td \Phi}=\frac{(1-\k^2\rho^2)(1+\k^2r^2)(\rho c_\zeta \, r c_\theta)^2}{P_4(\r,r,\z,\theta)} \ , 
\eea
where $P_4$   is expected to   have a polynomial dependence on $\rho$ and $r$
as well as a polynomial dependence on the trigonometric functions of $\zeta$ and $\theta$. 

%As in the lower-dimensional cases we expect the solutions with $C=0$ are special; 
%starting with a perturbative expansion in $\k$, we can find two such solutions for the dilaton:
Remarkably, as in the \adstr   case (cf. \rf{666},\rf{777}),   here we find two special solutions with $C=0$: 
\ba \label{a0Dil}
%\underline
{a=0}:& \ \ \ \ \ \ 
e^{-2\td \Phi}=\frac{(1-\k^2 \rho^2) ( 1+ \k^2  r^2)\, 
(\rho r c_\theta c_\zeta)^2}
{\big[1 -\k^2 ( \rho r )^2\big]^2\  \big[1-\k^2 ( \rho r s_\zeta s_\theta)^2\big]^2} \ , \ \ \qquad \qquad \qquad  \qquad \qquad \qquad  \ \, C=0 \ , \\
\label{a1Dil}
%\underline
{a=1}:& \ \ \   \ \ \  e^{-2\td \Phi}= \frac{(1- \k^2 \rho^2) (1+ \k^2 r^2)\,   (  \rho rc_\zeta c_\theta)^2}
{\big[1+ \k^2 r^2 -  \k^2(\rho s_\zeta)^2(1- r^2)\big]^2 \ \big[1-\k^2\rho^2 + \k^2(rs_\theta)^2(1+\rho^2)\big]^2}\ , \quad    C=0 \ . 
\end{align}
Undoing the T-duality, in  the original frame \rf{41},\rf{411}  the expressions for the   dilaton become 
%(we use the same  $\Phi$ to denote the original-frame dilaton) 
\ba 
\label{410}
{a=0}:& \ \ \ \ \ \   e^{-2\Phi}= \frac{(1- \k^2 \rho^2) (1+ \k^2 r^2)    (1 + \k^2 \rho^2 \bar \rho^2 )(1+ \k^2  r^2 \bar r^2)      }
{\big[1 -\k^2 ( \rho r)^2\big]^2\ \big[1-\k^2  ({\bar\rho}{\bar r})^2\big]^2}\ ,\qquad \bar \rho \equiv  \rho s_\zeta,\ \ \  \  \bar r\equiv  r s_\theta\ , 
\\
\label{4111}
{a=1}:& \ \ \ \ \ \  e^{-2\Phi}= \frac{(1- \k^2 \rho^2) (1+ \k^2 r^2)    (1+\k^2 \rho^2 \bar \rho^2 )(1+ \k^2  r^2 \bar r^2)      }
{\big[1+ \k^2 (r^2 -  \bar \rho ^2  +   \bar \rho ^2    r^2)\big]^2 \ \big[1+\k^2( \bar r^2 -  \rho^2 +   \rho^2    \bar r^2   )\big]^2}\ .
\end{align}
%where $\bar \rho = \rho s_\zeta, \  \bar r= r s_\theta $. 
%The identification with the $a=0$ and $a=1$ solutions in 
%the lower-dimensional cases is related to whether or not the function $P_4$ depends only on the product~$(r \rho)$.
%To compare   with   with the  known limits of the
%deformed  AdS$_5\times $S$^5$ supercoset background    
 Let us  now   consider  the 
$\kappa\to \infty$ and the $\k\to i$  % (pp-wave)
 limits   \ci{HRT}  of the above  expressions for the dilaton:
\begin{enumerate}[ $\k=\infty$: ]
\item As  discussed in \cite{HRT},  in the  $\k\to \infty$ limit 
%AAT
  the   natural    solution  for  the  dilaton   is expected to 
 be a product  of factors
depending separately on the AdS$_5$ and $S^5$ coordinates.
 Taking    $\k\to \infty$   in \rf{410}  we indeed  find   a factorization\foot{The  negative sign  may  be compensated by 
 a formal imaginary  constant shift  of $\Phi$.}
\be\label{125mn}
{a=0}: \ \ \qquad \qquad  \qquad  e^{-2\Phi}\Big|_{\k \to \infty}  \rightarrow -\frac{1}{(\rho r s_\zeta s_\theta)^2} \ , 
\ee  
%which agrees   with the expression for the dilaton found in \cite{HRT}(the ). 
  %in this limit (see eq.~(2.13) there).
%{\em -- up to a sign, implying need for imaginary shift of the dilaton? May be indication of some problem. }
At the same time,   the limit of the  $a=1$ expression \rf{4111}   does not factorize  for  $\k\rightarrow \infty$.
%indicating that  if  the two solutions  can be  embedded  into a family  parametrized   by 
 %$a$,  one should  choose  $a=a(\k)$ so that $a(\k \to \infty)=0$.   %  must vanish in the limit. 
  \end{enumerate}
\begin{enumerate}[ $\k=i$: ]
\item In the $\k\to i$  limit the dilaton  may also  be expected to    factorize  \cite{HRT}.\foot{The explicit form of the full ``pp-wave" 
background  corresponding to the $\k=i$ limit of the deformed \adss   solution was not found in \ci{HRT}.}
 However, this does  not happen 
for  the $a=0$  expression \rf{410}. % indicating 
%$a=0$ expression \eqref{aeq0ads5}, 
%\be \label{a0k0} e^{-2\Phi^{(0)}}\Big|_{\k=i}=\frac{[1+\rho^2][1-r^2]  [\rho r c_\theta c_\zeta]^2} {[1 +(r\rho)^2]^2[1+(\rho r s_\zeta s_\theta)^2]^2} \ , \ee 
%that  one   should   have $a(\k\to i) \ne 0$. 
%
%This expression  does not seem to have any simple interpretation (in particular, it does not factorize) -- we had a similar problem 
%with $a=0$ solutions in lower dimensions. 
At the same time,  the  $\k\to i$  limit of  the $a=1$ dilaton \rf{4111}    does    factorize
\bea\la{413} 
a=1:\qquad \qquad 
e^{-2\Phi} \Big|_{\k \to i }   \rightarrow  \frac{  (1- \rho^4s_\zeta^2)(1- r^4s_\theta^2)   }{(1+\rho^2) (1-r^2)[1+(\rho s_\zeta)^2]^2[1-(r s_\theta)^2]^2} \ . 
\eea
%This expression factorizes between AdS$_5$ and S$^5$, as expected from the analysis of \cite{HRT}. 
\end{enumerate}
Thus, as in the lower-dimensional cases, %different limits of the supercoset geometry require different values (an analog) of 
%the parameter $a$ -- 
%consistency  with the limits of the  supercoset model  appears to
 it seems natural to expect   the existence of a one-parameter  family of solutions 
with  $a=a(\k)$ chosen  so that  $a(  i)=1$ and $a( \infty)=0$.
%-- and thus an interpolating solution for the three scalar equations is necessary  to reproduce the supercoset data.

%gnoring for the moment any comparison with supercoset information, 

At the same time,   it is possible to check that the algebraic Rainich
conditions \eqref{10dRainich}  for existence of the $F_5$   flux 
are not satisfied by the stress tensor containing the contribution 
of the  T-dual frame dilatons eqs.~\eqref{a0Dil} and \eqref{a1Dil} only. 
This  indicates  that  one should  look for more general  solutions with several  RR fields 
excited. 
%that these expressions are not consistent with the assumption that they are supported only by a five-form field strength.
%
This  is analogous to our   earlier observation that the  $a$-family  of scalar  field solutions in the   deformed 
%$a=0,1$ 
AdS$_3\times$S$^3$ case   \rf{3666},\rf{366}
 cannot  be supported by  just one   3-form RR  field strength.

%%%%%%%%%%%%%%%%%%%%%%%%%%%
\section{Some properties of  the deformed backgrounds}
%%%%%%%%%%%%%%%%%%%%%%%%%%%%%%%

While  we did not find the  full   solution in the  deformed  \adss case 
some of its properties    are already  evident from the form of the metric and the dilaton
and are  shared    with    the corresponding \adst and \adstr  solutions. 
All these   deformed  backgrounds   represent a novel class of 
non-supersymmetric  type IIB supergravity solutions   which   have factorized  string-frame metric but non-factorized 
dilaton and RR fields.

 For all  the three  deformed  string-frame metrics  \rf{22},\rf{31},\rf{41}    in dimensions 4, 6 and 10
  there is a (naked)  curvature singularity at $\rho =\k^{-1}$  (cf. \rf{cu}). 
  The integrability  of the underlying sigma models \ci{foz,fat,luk,DMV}
  implies, in particular,    that it should   be possible to  find the  explicit form of the corresponding geodesics  
   and study  their approach to the singularity.  
  At the same time, %the   integrable structure  of the string sigma model may be suggesting that 
  concentrating on the point-like limit may be misleading: 
  one  may  need to    investigate   if the    string probes  ``see"   the singularity. For
   example, attempting to probe it with a long spinning 
folded string shows that the fold-points of the string remain at some finite distance
 from the singularity \cite{fhr,Kameyama:2014vma}.\foot{For a discussion of  classical string solutions in deformed  geometry  
 see also \ci{Arutyunov:2014cda,Banerjee:2014bca,Ahn:2014aqa}.}
%If this situation were to persist with other string probes it would suggest that it may not be necessary to resolve this singularity.
%While our attempt to complete the deformed AdS$_5\times$S$^5$ metric to a solution of type IIB supergravity was 
%unsuccessful,  we may comment on some of the physical properties of the solution.  

Since the string-frame dilaton equation  is independent of the  RR fluxes, the singularities of the dilaton  are 
determined by the singularities of the metric and  the NSNS $B$-field. 
From the dilaton equation  (or the exact solutions \rf{999},\rf{3666},\rf{316},\rf{72},\rf{410},\rf{4111})
one concludes that  near this point 
$ e^{ \Phi } \to   (1 - \k^2 \rho^2)^{-1/2} \to \infty$. This means that the effective  string coupling blows up, 
  suggesting that  one cannot study the near-singularity 
region using   string perturbation theory.\foot{From the  10d type IIB  supergravity perspective, 
 one may pass to the S-dual frame,  where the dilaton is small near  $ \rho = \k^{-1}$. 
However, the metric will continue to be singular. % and some mechanism must be invoqued for resolving the singularity.
Also, the S-dual  solution  will no longer  have   a deformed supercoset background  interpretation.} 

This conclusion may, however, be premature:
%This  suggests  that  the  curvature singularity  may  not within the realm of  string perturbation theory. 
%It may however also suggest that its  existence is itself an artifact of string perturbation theory. 
due to   lack of supersymmetry   the leading-order  supergravity solution may  receive 
non-trivial $\alpha'$  corrections  that may smear the singularity out in both the metric  and the dilaton. 
Clarifying  this issue   requires a better understanding  of the  underlying 
deformed supercoset   model at the quantum level.  

%If $\kappa$ is commensurate with $\alpha'$ perturbation theory should also be resummed before the singularity 
%of the metric or of the dilaton can be analyzed.
%Nevertheless, given a solution of supergravity equations of motion, one may nevertheless attempt to study it as such,
%without attempting to guess what non-perturbative effects might correct the classical picture.
%(ii) Regardless of the frame, the metric possesses  a naked singularity at $\kappa\rho=1$. It is not a priori clear how this
%singularity is resolved or whether it should be resolved at all. For example, attempting to probe it with a long spinning 
%folded string shows that the endpoints of the string remain at some finite distance from the singularity \cite{fhr, jap}.
%If this situation were to persist with other string probes it would suggest that it may not be necessary to resolve this singularity.

 It is interesting to note that while both the  deformed metric  $g_{mn}$ and the dilaton  $\Phi$ 
 are singular, in all AdS$_n\times$S$^n$ cases
the  ``T-duality invariant" volume density  $e^{-2\Phi} \sqrt {-g} $ is regular at $  \rho=\k^{-1}$. 
For example, if one performs  a  formal T-duality along the time $t$ direction, in, e.g.,  \rf{22} 
one gets  a regular 
metric with a horizon at  $ \rho=\k^{-1}$ and   with the   T-dual  string  coupling  $e^\Phi$   vanishing  at that point.  
While this time-like T-duality is   a formal transformation (the resulting type IIA background will  have complex fluxes)
% the presence of the horizon and the regularity of the dilaton 
this may be suggesting  a  hidden  regularity of  the original  type IIB background.\foot{One may draw an  analogy 
 with the T-duality in flat 2-space in polar coordinates or in Rindler space:  
a   background $ds^2 = - r^{-2} dt^2 + dr^2, \ \ \Phi= -\ln r$ with curvature and dilaton singularity at $r=0$   is T-dual to a regular 
 one   with $ds^2 = - r^{2} dt^2 + dr^2, \ \ \Phi= 0$.}

\iffalse
As in $AdS_3 \times S^3$ case one  may expect that \rf{a0}  and \rf{a1}  exhaust all possible 
solutions   with $C=0$. But where is analogy with $AdS_2 \times S^2$ -- there  for $a=1$ dilaton we needed  $C\not=0$?
There seems to be rather analogy with $AdS_3 \times S^3$ case where there was no $C$, so this may be ok. 
{\bf Once again, we need  $a(\k)$ family of solutions  to satisfy both $\k=i$ and $\k=\infty$ limits}
\fi

%%%%%%%%%%%%%%%%%%%%%%%%%%%%%%%%%%%%%%%%%%%%%%%%%%%%%
\section*{Acknowledgements}
We would like to thank B. Hoare  for  very useful discussions and  also  D. Marolf and W. Kelly for pointing out the 
references  \cite{Torre:2012hw,Torre:2013nia}. OL and RR thank the Theoretical Physics Group at Imperial College for hospitality, and RR 
acknowledges the  hospitality of the KITP at UCSB. OL is supported by the NSF grant PHY-1316184.
The work of RR was supported in part by the US Department of Energy under contract DE-SC0008745 
and while at KITP also by the National Science Foundation under Grant No. NSF PHY11-25915.
The work of AAT was supported by the  STFC grant ST/J000353/1,
 by the ERC Advanced grant No.290456 and also by the RSF grant 14-42-00047.  

%%%%%%%%%%%%%%%%%%%%%%%%%%%%%%%%%%%%%%%%%%%%

%\newpage

\appendix

\section{Equations of motion and embedding into 10d supergravity}
\label{AppEOM}

\renewcommand{\theequation}{A.\arabic{equation}}
\setcounter{equation}{0}

In sections \ref{Sec4D} and \ref{Sec6D} we discussed  $d=4$ and $d=6$ supergravities truncated to
 two scalar fields (the dilaton and the RR scalar) and one $d/2$-form field. The
  corresponding actions  may be written as 
\bea\label{UnivAaction}
S=\int d^d x\sqrt{-g}\left[e^{-2\Phi}\big(R+4(\del\Phi)^2\big)-
\frac{e_2}{4}F_{mn}F^{mn}-\frac{e_3}{12}
{ F}_{mnp}{ F}^{mnp}-\frac{1}{2}(\del C)^2\right] \ , 
\eea where  the coefficients $(e_2,e_3)$  are
\beaa\label{ValuesOfe}
d=4: \ \ \ e_2=1,\quad e_3=0;   \ \ \ \ \ \ \ \ 
d=6:  \ \ \ e_2=0,\quad e_3=1 \ .
\eeaa
The equations of motion coming from this action  are 
\ba
\label{one}
&e^{-2\Phi}R_{mn}=-2e^{-2\Phi}\nabla_m\nabla_n\Phi+
\frac{e_3}{4}\Big({ F}_{mpq}{ F_n}^{pq}-\frac{1}{6}g_{mn}{F}_{spq}{F}^{spq}\Big)\nonumber\\
&\qquad\qquad\quad
+\frac{e_2}{2}\Big(F_{mk}{F_n}^k-\frac{1}{4}g_{mn}F^2\Big) +\frac{1}{2}\Big[\d_m C\d_n C-\frac{1}{2}g_{mn}(\d C)^2\Big]\ , \\
\label{onePrime}
&  \nabla_m F^{mn} =0 \ , \qquad     \nabla_m F^{mnk}  =0 \ , \\ 
&\Big( -\nabla^2  +  {1 \ov 4 } R\Big) e^{-\Phi}=0\ , \ \ \ \ \qquad     \nabla^2 C=0 \ . \la{a5}
\end{align}
It is convenient to separate the trace of the Einstein  equation and combine it with the other scalar equations.
Using the fact that the trace of the stress tensor of the $d/2$-form in $d$ dimensions  vanishes,   we have from \rf{one} 
%%%%%
%%%%%
\bea
\label{two}
e^{-2\Phi}R=-2e^{-2\Phi}\nabla^2\Phi - \frac{d-2}{4}\, \d_m C\d^m C \ .
\eea
Then \rf{a5} with \rf{two}   give 
\bea
\nabla^2 \Big( e^{-2 \Phi}+\frac{d-2}{8}C^2\Big) =0 \ ,
\label{cc}
\eea
which may be used in place of any of the three scalar equations in \rf{a5} and \rf{two}. 

In  the 10d case     with non-vanishing  $B$-field   we get   the following   forms   of the scalar equations 
  \ba %\lab{eqs}
  &R - {1 \ov 4} H^2 + 2 \nabla^2 \Phi +    2 e^{2\Phi}( \partial  C)^2 =0, \ \ \ \qquad 
  \Big[ -\nabla^2  +  {1 \ov 4 } \big( R  -  { 1 \ov 12} H^2 \big) \Big]e^{-\Phi}=0\ , \\
 % R- { 1 \ov 12} H^2   + 4 \nabla^2 \phi - 4 (\partial \phi)^2 =0 \ , \ \ \ \ \ \ \     \\ 
    %\textstyle  
 &  \nabla^2  \Big( e^{-2 \Phi}   +  C^2\Big)  =  { 1 \ov 6}   H^2 \ , \ \ \ \ \qquad \qquad\qquad     \nabla^2 C =0 \  .
\end{align}
Let us  now review the embedding of the four-- and 
six--dimensional systems (\ref{UnivAaction}), (\ref{ValuesOfe}) in 10D supergravity.

The undeformed AdS$_3\times$S$^3$ solution can be embedded in type IIB supegravity by identifying $F_3$ in (\ref{UnivAaction}) with RR 3-form  field strength  in ten dimensions. To embed the deformed 6d solution, we also  identify $C$ with the RR scalar  in ten dimensions, i.e. the starting point is the following  truncated 10d action 
\bea
S=\int d^{10} x\sqrt{-g_{10}}\left[e^{-2\Phi}\big(R+4(\d\Phi)^2\big)-\frac{1}{12}F_{MNP}F^{MNP}-
\frac{1}{2}(\d C)^2\right].
\eea
To perform the reduction to 6d, we write the ten--dimensional metric as\footnote{It is easy to check that more general warp factors on the torus, i.e.  $\sum e^{A_i} dy_idy_i$, do not  lead to additional constraints for solutions with $A_i=A_j$, so that  we may focus only on the volume mode.}
\bea
ds_{10}^2=g_{mn}dx^mdx^n+e^A dy_idy_i\,,
\eea
where $y_i$ are flat coordinates on $T^4$. 
The standard dimensional reduction on the 4-torus then gives (see, e.g.,  \cite{sen}) 
\bea
S=\int d^{6} x\sqrt{-g}\Big[e^{-2(\Phi-A)}\Big(R+
4\big[\d(\Phi-A)\big]^2-
(\d A)^2\Big)-\frac{e^{2A}}{12}F_{mnp}F^{mnp}-
\frac{e^{2A}}{2}(\d C)^2\Big] \,.
\eea
This reduces to (\ref{UnivAaction}) for $A=0$, but equation of motion for $A$ leads to an additional constraint:
\bea\label{Constr}
\frac{1}{12}F_{mnp}F^{mnp}+\frac{1}{2}(\d C)^2=0 \ . 
\eea
This relation is satisfied by (\ref{316}), (\ref{72}), (\ref{32260}), (\ref{3226}). 

\ 

The undeformed AdS$_2\times$S$^2$ solution can be embedded in type II  10d supergravity
 in two different ways \cite{STWZ}, which are related by T-dualities. 
 %We begin with discussing embedding the deformation in IIA SUGRA. 
 In the absence of the Kalb--Ramond field, the action for type IIA  supergravity is 
\bea
S=\int d^{10} x\sqrt{-g_{10}}\Big(e^{-2\Phi}\big[R+4(\d\Phi)^2\big]-\frac{1}{48}F_{MNPQ}F^{MNPQ}-
\frac{1}{4}F_{MN}F^{MN}\Big)\ . 
\eea
Choosing  the ansatz  ($z_i$ are 3 complex coordinates of 6-torus) 
\beaa\label{Metr10D6red}
ds_{10}^2&=&g_{mn}dx^mdx^n+e^A dz_id{\bar z}_i
 \ , \qquad \qquad F^{(2)}=\frac{1}{2\sqrt{2}}{\tilde F}_{mn}dx^m \wedge dx^n\ , \nonumber  \\
F^{(4)}&=&
\frac{1}{2\sqrt{2}}{F}_{mn}dx^m \wedge dx^n\wedge J_2+
\frac{1}{2}dC\wedge \mbox{Re}  \, \Omega_3 \  , \\
J_2 &\equiv& \frac{i}{2}dz_k\wedge d{\bar z}_k  \ ,\qquad \qquad  \ \  \Omega_3 \equiv  dz_1\we dz_2\we dz_3\,, \ \ \ \nonumber
\eeaa
and reducing on the 6-torus we find
\ba \nonumber 
S= &\int d^{4} x\sqrt{-g} \Big[ e^{-2\Phi+3A}\Big(R+
4\big[\d(\Phi-\frac{3}{2}A)\big]^2-\frac{3}{2}
(\d A)^2\Big) \\  & \qquad\qquad\qquad   -   \frac{3e^{A}}{8}F_{mn}F^{mn}-  \frac{e^{3A}}{8}{\tilde F}_{mn}
{\tilde F}^{mn}
- 
\frac{1}{2}(\d C)^2\Big]\,.   \label{ReducedIIA6D}
\end{align}
To have a solution with $A=0$, we must set 
\bea
F_{mn}F^{mn}+{\tilde F}_{mn}{\tilde F}^{mn}=0\,,
\eea
and this constraint can be satisfied by imposing a relation
\bea\label{StarF1110}
{\tilde F}=\star  F\,.
\eea
Substituting this relation for ${\tilde F}$ into (\ref{ReducedIIA6D}) and setting $A=0$, we recover (\ref{UnivAaction}) with $e_2=1$, $e_3=0$.
 
 %v3
The deformed AdS$_2\times$S$^2$  solution  can  be also embedded into type 
IIB theory as
\beaa\label{EmbedIIB}
ds_{10}^2&=&g_{mn}dx^mdx^n+e^A dz_id{\bar z}_i\,,
\nonumber\\
F^{(3)}&=&\frac{1}{2}dC\wedge J_2+\frac{1}{12}\star\big(dC\wedge J_2 \we J_2 \we J_2    \big)\,,  \\
F^{(5)}&=&\frac{1}{2}F \wedge \mbox{Im}\,  \Omega_3 -
\frac{1}{2} \star_4({F}) \wedge \mbox{Re} \,  \Omega_3 
\ , \qquad \qquad F\equiv \frac{1}{2}F_{mn}dx^m \we dx^n \ .   \nonumber
\eeaa
To write the action we relax the self-duality conditions by replacing $F^{(5)}$ in \rf{EmbedIIB}  with 
\bea
F^{(5)}=\frac{1}{\sqrt{2}}F\wedge \mbox{Im}\,  \Omega_3\,
\eea
that leads to equivalent equations of motion. 
%which solves the same equations of motion.
 The dimensional reduction of the type IIB action
\beaa
S=\int d^{10} x\sqrt{-g_{10}}\Big[e^{-2\Phi}(R+4(\d\Phi)^2)-\frac{1}{12}  F_{MNK} F^{MNK} -\frac{1}{480}  F_{MNKLP} F^{MNKLP}\Big]\ \,,
\eeaa
then gives
\beaa\label{4DactionFromD3}
S&=&\int d^{4} x\sqrt{-g}\Big[ e^{-2\Phi+3A}\Big(  R+
4\big[\d(\Phi-\frac{3}{2}A)\big]^2-\frac{3}{2}
(\d A)^2  \Big) \nonumber\\
&&\qquad \qquad \qquad  -   \frac{1}{4}F_{mn}F^{mn} - 
\frac{3e^A}{8}(\d C)^2-  \frac{e^{-3A}}{8}(\d C)^2\Big]\,.
\eeaa
This coincides with (\ref{UnivAaction}) for configurations with $A=0$, and the equation of motion for $A$ does not introduce additional constraints.
%%The solution  for $F_{mn}$ is to be chosen so that $F_5$ in \rf{Metr10D6red}  becomes self-dual. 

To summarize, we have demonstrated that the reduction of  type II  10d  supergravity 
 reproduces the 4d action (\ref{UnivAaction}), but its 6d counterpart  must be supplemented by the constraint (\ref{Constr}). 

%%%%%%%%%%%%%%%%%%%%%%
\section{Rainich conditions in four dimensions}

\renewcommand{\theequation}{B.\arabic{equation}}
\setcounter{equation}{0}
\label{AppRainich}
As discussed in section 3, 
to test whether a given stress--energy tensor can be sourced 
by a particular type of flux, we need a 
generalization of  the Rainich condition to higher dimensions.
 To review  the original condition in 4d,  let us start with Maxwell stress tensor 
\bea
{T_m}^n=F_{mk}F^{kn}-\frac{1}{4}\delta_m^n F_{sk}F^{ks} \ , 
\eea
%This tensor satisfies two algebraic Rainich conditions:
which satisfies  the two algebraic conditions 
\bea\label{OrigRainich}
{T_m}^m=0,\qquad \qquad {T_m}^k{T_k}^n=
\frac{1}{4}\delta_m^n {T_s}^k{T_k}^s \ . 
\eea 
The first condition is obvious,  while  to prove the second one, 
we can go to the orthonormal frame and perform a (coordinate--dependent) rotation
 to put $F_{mn} $ into  a  blog-diagonal  form
\bea
{F_m}^k=\left(
\begin{array}{cccc}
0&a_1&0&0\\
-a_1&0&0&0\\
0&0&0&a_2\\
0&0&-a_2&0
\end{array}
\right) \ ,  \ \ \ \ \ \ \ \ \ \ \ \  {T_m}^n=\frac{1}{2}(a_2^2-a_1^2)\,\mbox{diag}(1,1,-1,-1) \ . 
\eea
%Then\bea\label{4DRainStress}\eea and algebraic Rainich relations follow. 
It us useful to note that   the Rainich conditions  \rf{OrigRainich}  imply that\foot{Here $T^3$ stands for $T_n^k T_k^m T_m^r$, etc.}
\bea\label{ModRainich}
\mbox{tr}\,T=0,\qquad\qquad  \mbox{tr}\,T^3=0 \ .
\eea 
%To demonstrate that these conditions are equivalent to (\ref{OrigRainich}), 
Indeed,    a $4\times 4$ matrix $T$ satisfies its own characteristic equation:
\bea
T^4-\frac{1}{2}\mbox{tr}(T^2)\,T^2-
\frac{1}{3}\mbox{tr}(T^3)\,T+\mbox{det}(T)=0 \ , 
\eea
 where we used that $\tr\, T=0$. 
 Then using  (\ref{OrigRainich})   we  conclude that $\mbox{tr}(T^3)\,T=0$, implying  (\ref{ModRainich}).

    %we used the tracelessness of $T$, which is a part of both (\ref{OrigRainich}) and (\ref{ModRainich}). Substitution of (\ref{OrigRainich}) into $T^2$ and $T^4=(T^2)^2$ in the last relation leads to equation \bea \mbox{tr}(T^3)\,T=0, \eea
%which implies (\ref{ModRainich}). Notice that that restriction 
%(\ref{OrigRainich}) on $T^2$ seems stronger than (\ref{ModRainich}), which only implies a relation for $T^4$. 

\end{document}